\documentclass[submission, Phys]{SciPost}

\pdfoutput=1

\usepackage[utf8]{inputenc} 
\usepackage[T1]{fontenc} 	
\usepackage[english]{babel} 

\usepackage[bitstream-charter]{mathdesign}
\usepackage{geometry} 		
\usepackage{amsmath} 		
\usepackage{mathtools} 		
\usepackage{float} 			
\usepackage{graphicx} 		
\usepackage{tabularx} 		
\usepackage{booktabs} 		
\usepackage{color, xcolor} 	
\usepackage{pdfpages} 		
\usepackage{extarrows} 		
\usepackage{multirow} 		
\usepackage{multicol} 		
\usepackage{caption} 		
\usepackage{subcaption} 	
\usepackage{enumitem} 		
\usepackage{xspace} 		
\usepackage{stackrel} 		
\usepackage{tikz} 			
\usetikzlibrary{calc}
\usepackage{braket} 		
\usepackage{bm} 			
\usepackage{tensor} 		
\usepackage{slashed} 		
\usepackage{siunitx} 
\usepackage{lastpage} 		
\usepackage{cite} 			
\usepackage[normalem]{ulem} 
\usepackage{fontawesome} 	
\usepackage{tocloft} 		
\usepackage{titlesec} 		
\usepackage{doi} 			
\usepackage[most]{tcolorbox} 					
\usepackage[capitalize]{cleveref} 	
\usepackage[nottoc, notlot, notlof]{tocbibind} 	
\usepackage[ruled, vlined]{algorithm2e} 		
\usepackage{makecell}
\usepackage{pifont}
\usepackage{stackrel}
\usepackage{placeins}

\hypersetup{
	pdftitle={Generative Unfolding of Jets and Their Substructure},
	pdfauthor={Antoine Petitjean, Anja Butter, Kevin Greif, Sofia Palacios Schweitzer, Tilman Plehn, Jonas Spinner, Daniel Whiteson},
    linktoc=section,
	breaklinks=true,			
	colorlinks=true, 			
	linkcolor={red!50!black}, 	
	citecolor={blue!50!black}, 	
	urlcolor={blue!80!black} 	
} 

\DeclareSymbolFont{usualmathcal}{OMS}{cmsy}{m}{n}
\DeclareSymbolFontAlphabet{\mathcal}{usualmathcal}

\SetArgSty{textnormal}
\SetKwComment{Comment}{{\small\#}~}{}
\SetCommentSty{mycommfont}

\setitemize{itemsep=2pt, parsep=0pt} 				
\setenumerate{itemsep=2pt, parsep=0pt} 				
\crefname{section}{Sec.}{Secs.}
\crefname{equation}{Eq.}{Eqs.}
\crefname{figure}{Fig.}{Figs.}
\crefname{table}{Tab.}{Tabs.}
\Crefname{section}{Section}{Sections}
\Crefname{equation}{Equation}{Equations}
\Crefname{figure}{Figure}{Figures}
\Crefname{table}{Table}{Tables}

\usetikzlibrary{arrows, shapes.geometric, arrows.meta, shapes, decorations.pathreplacing, fit, patterns, patterns.meta}

\definecolor{Rcolor}{HTML}{E99595}
\definecolor{Gcolor}{HTML}{C5E0B4}
\definecolor{Bcolor}{HTML}{9DC3E6}
\definecolor{Ycolor}{HTML}{FFE699}

\tikzstyle{expr} = [circle, minimum width=1.8cm, minimum height=1.8cm, text centered, align=center, inner sep=0, draw,font=\LARGE]
\tikzstyle{txt_huge} = [align=center, font=\Huge, scale=2]
\tikzstyle{txt} = [align=center, font=\LARGE]
\tikzstyle{cinn} = [double arrow, double arrow head extend=0cm, double arrow tip angle=130, shape border rotate=90, inner sep=0, align=center, minimum width=2.1cm, minimum height=2.3cm, fill=Bcolor, draw,font=\LARGE]
\tikzstyle{cinn_black} = [cinn, minimum height=2.5cm, fill=black]
\tikzstyle{arrow} = [thick,-{Latex[scale=1.0]}, line width=0.2mm, color=black]
\tikzstyle{line} = [thick, line width=0.2mm, color=black]
\tikzstyle{loss} = [rectangle, align=center,  minimum width=1.8cm, minimum height=1.5cm,fill=Rcolor,font=\LARGE, rounded corners]
\tikzstyle{xt} = [rectangle, align=center,  minimum width=4cm, minimum height=1.5cm,fill=Gcolor,font=\Large, rounded corners]
\tikzstyle{xts} = [rectangle, align=center,  minimum width=1cm, minimum height=1.5cm,fill=Gcolor,font=\Large, rounded corners]

\definecolor{gatr-green}{HTML}{419108}

\newcommand{\qqquad}{\qquad\quad}

\newcommand{\loss}{\mathcal{L}} 

\newcommand{\punf}{p_\text{unfold}}
\newcommand{\psim}{p_\text{sim}}

\newcommand{\pd}{p_\text{data}}

\newcommand{\xp}{x_\text{part}}
\newcommand{\xr}{x_\text{reco}}

\newcommand{\fastjet}{\textsc{FastJet}\xspace}
\newcommand{\pythia}{\textsc{Pythia}\xspace}

\newcommand{\delphes}{\textsc{Delphes}\xspace}

\newcommand{\madgraph}{\textsc{Madgraph}\xspace}

\newcommand{\arXiv}[2][]{%
\ifthenelse{\equal{#1}{}}%
{\href{http://arxiv.org/abs/#2}{arXiv:#2}}%
{\href{http://arxiv.org/abs/#2}{arXiv:#2~[#1]}}}

\newcommand{\gev}{\text{GeV}}

\def\slashchar#1{\setbox0=\hbox{$#1$} 
\dimen0=\wd0 
\setbox1=\hbox{/} \dimen1=\wd1 
\ifdim\dimen0>\dimen1 
\rlap{\hbox to \dimen0{\hfil/\hfil}} 
#1 
\else 
\rlap{\hbox to \dimen1{\hfil$#1$\hfil}} 
/ 
\fi}

\newcommand{\tikznode}[2]{%
\ifmmode%
\tikz[remember picture,baseline=(#1.base),inner sep=0pt]{\node (#1) {$#2$};}%
\else \tikz[remember picture,baseline=(#1.base),inner sep=0pt]{\node (#1) {#2};}%
\fi}

\def\mathswitchr#1{\relax\ifmmode{\text{#1}}\else$\text{#1}$\xspace\fi} \def\mathswitch#1{\relax\ifmmode#1\else$#1$\xspace\fi}

\newcommand{\Npart}{\mathswitch{N_\text{part}}}
\newcommand{\Ndet}{\mathswitch{N_\text{reco}}}
\newcommand{\Jpart}{\mathswitch{J_\text{part}}}
\newcommand{\Jdet}{\mathswitch{J_\text{reco}}}

\graphicspath{{./figs/}}

\begin{document}

\begin{center}
    {\Large \textbf{Generative Unfolding of Jets and Their Substructure}}
\end{center}

\begin{center}
    Antoine Petitjean\textsuperscript{1}, 
    Anja Butter\textsuperscript{1,2},
    Kevin Greif\textsuperscript{3}, \\
    Sofia Palacios Schweitzer\textsuperscript{1,4}, 
    Tilman Plehn\textsuperscript{1,5}, 
    Jonas Spinner\textsuperscript{1,6}, and
    Daniel Whiteson\textsuperscript{3}
\end{center}

\begin{center}
    {\bf 1} Institute for Theoretical Physics, Universit\"at Heidelberg, Germany \\
    {\bf 2} LPNHE, Sorbonne Universit\'e, Universit\'e Paris Cit\'e, CNRS/IN2P3, Paris, France \\
    {\bf 3} Department of Physics and Astronomy, University of California, Irvine, California, USA \\
    {\bf 4} NHETC, Department of Physics \& Astronomy, Rutgers University, Piscataway, NJ, USA \\
    {\bf 5}  Interdisciplinary Center for Scientific Computing (IWR), Universit\"at Heidelberg, Germany \\
    {\bf 6}  Institute for Particle Physics Phenomenology, Department of Physics, Durham University, UK
\end{center}

\begin{center}
    \today
\end{center}

\section*{Abstract}
{\bf Unfolding, for example of distortions imparted by detectors, provides suitable and publishable representations of LHC data. Many methods for unbinned and high-dimensional unfolding using machine learning have been proposed, but no generative method scales to the several hundred dimensions necessary to fully characterize LHC collisions. This paper proposes a 3-stage generative unfolding framework that is capable of unfolding several hundred dimensions. It is effective to unfold the jet-level kinematics as well as the full substructure of light-flavor jets and of top jets, and is the first generative unfolding study to achieve high precision on high-dimensional jet substructure.}


\vspace{10pt}
\noindent
\rule{\textwidth}{1pt}
\tableofcontents
\thispagestyle{fancy}
\noindent
\rule{\textwidth}{1pt}
\vspace{10pt}

\clearpage
\section{Introduction}
\label{sec:intro}

Experiments at the LHC observe particle interactions at high energy to study the nature of matter and forces. However, detectors have finite resolution and acceptance. Measuring physical parameters requires including these detector effects in predictions, which can be compared to data. Because such effects are expensive to model and require access to proprietary software, an alternative is the reverse, to remove the impact of detector, \textit{unfolding} the data, which allows direct comparison with theoretical predictions and across experiments. The measured spectra are defined at \textit{particle level}, the phase space of the stable particles before detector effects are modeled. Traditional unfolding methods require binning of the data and are limited to a small number of dimensions~\cite{Cowan:2002in}. Machine-learning (ML)-based~\cite{Plehn:2022ftl} unfolding techniques have overcome these limitations, offering un-binned unfolding in many simultaneous dimensions. Two classes of ML unfolding have emerged, discriminative and generative.

Discriminative unfolding algorithms train classifiers to re-weight simulated events at particle level to produce un-binned unfolded spectra~\cite{Andreassen:2019cjw,Chan:2023tbf,Desai:2024yft,Zhu:2024drd,Falcao:2025jom,Canelli:2025ybb} and have recently been used in several measurements~\cite{H1:2021wkz, H1:2023fzk, H1:2024mox, LHCb:2022rky, Huang:2025ziq, Pani:2024mgy, Song:2023sxb, ATLAS:2024rpl, ATLAS:2025qtv, CMS:2025sws, Badea:2025wzd}. They assume that the sample of simulated events at particle level has sufficient support to construct the particle level data distribution. In this case, they are known to be accurate even in high- and variable-dimensional unfolding tasks, such as unfolding the entire substructure of high-multiplicity jets~\cite{Komiske:2022vxg}.

Generative algorithms learn a mapping between detector-level events and particle-level events, formalized as a posterior density conditional on observed detector-level events to allow sampling from the truth-level distributions~\cite{Datta:2018mwd,Bellagente:2019uyp,Bellagente:2020piv,Vandegar:2020yvw,Howard:2021pos,Backes:2022sph,Diefenbacher:2023wec,Ackerschott:2023nax,Shmakov:2023kjj,Shmakov:2024gkd, Pazos:2024nfe, Butter:2024vbx, Favaro:2025psi, Butter:2025via}.
Generative algorithms produce unweighted events and could have complementary strengths and weaknesses compared to discriminative methods.
However, they require solving the challenging generative ML task of modeling the full particle-level distribution. They have been shown to produce excellent accuracy for fixed-dimensional phase spaces of several dimensions~\cite{Huetsch:2024quz,Favaro:2025psi}, and have also shown promise for variable-dimensional spaces~\cite{Shmakov:2024gkd}. However, no generative unfolding algorithm has been shown to scale to high-dimensional unfolding.

Although many measurements only require a few dimensions for the final comparison, the inclusion of additional dimensions on which the detector response depends gives higher accuracy~\cite{Desai:2025mpy}. Many measurements, such as those using energy-energy correlators~\cite{Lee:2022uwt, Moult:2025nhu, Holguin:2024tkz} are best approached by unfolding every final state particle. Unfolding without dimensionality reduction also preserves the maximum amount of information for future analysis. If the unfolding is differential in the kinematics of every final-state particle, any observable can be easily measured post-hoc, allowing reinterpretation for unforeseen use cases.

This paper describes a new generative method capable of unfolding several hundred dimensions, allowing for unfolding of high-multiplicity jets. It makes use of conditional flow matching (CFM) generative networks~\cite{chen2018neural,lipman2023flowmatchinggenerativemodeling,albergo2023stochastic,Butter:2023fov} with and without a Lorentz-equivariant architecture~\cite{Brehmer:2024yqw,Favaro:2025pgz} to solve three generative tasks. First, a multiplicity prediction network provides an estimate of the number of particles contained in the particle-level jet. Second, the jet kinematics are predicted and used to provide a physically motivated rescaling of the kinematics of the jet constituents. Finally, the full-dimensional phase space of the jet constituents conditioned on the predicted multiplicity and the jet kinematics is generated. The method accurately unfolds the full substructure of jets in two benchmark datasets, one containing light quark and gluon initiated jets produced in association with a $Z$-boson, and one containing jets produced by the hadronic decay of boosted top quarks.

This paper is organized as follows.
Section~\ref{sec:gen_unfolding} provides an overview of generative unfolding and the conditional flow matching framework.
Section~\ref{sec:method} details the methods used for each step of the procedure.
Sections~\ref{sec:z_plus_jets} and~\ref{sec:ttbar} illustrate the performance of the method on light-jet and top-jet benchmarks, and conclusions are drawn in Section~\ref{sec:conclusion}.

\section{Generative Unfolding}
\label{sec:gen_unfolding}

\begin{figure}[t]
    \centering
    \input{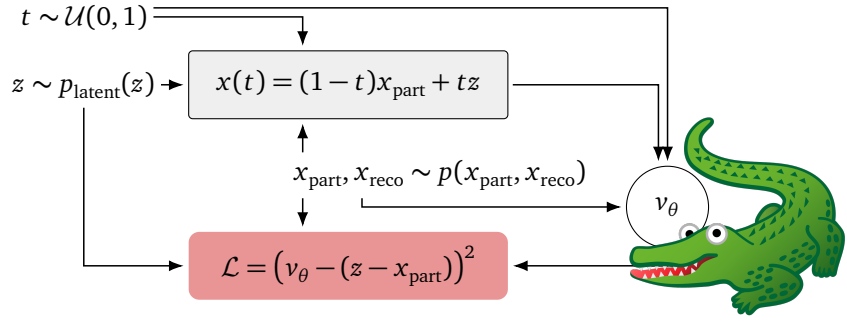}
    \caption{Visualization of the CFM training procedure for a generative unfolding task. The network output $v_\theta$ can be estimated by a neural network equipped with the L-GATr architecture. See text for details.}
    \label{fig:money}
\end{figure}

Given observed data $\xr$ and its phase space density $\pd(\xr)$, the goal of unfolding is to estimate the density $\punf(\xp)$ in the unobserved parton-level or particle-level phase space $\xp$.  This requires a convolution with conditional $p(\xp|\xr)$, 
\begin{align}
    \punf(\xp) = \int d\xr \, p(\xp | \xr) \pd(\xr) \; ,
    \label{eq:gen_unfolding1}
\end{align}
which is generally unavailable. However, the related conditional $p(\xr|\xp)$ is implicitly encoded in the forward simulation chain, transforming the density $\psim(\xp)$ into $\psim(\xr)$.
\begin{alignat}{9}
  & \psim(\xp)
  \quad \xleftrightarrow{\text{unfolding inference}} \quad 
  && \punf(\xp)
  \notag \\
  & \hspace*{-9mm} {\scriptstyle p(\xr|\xp)} \Bigg\downarrow
  && \hspace*{+6mm} \Bigg\uparrow {\scriptstyle p(\xp|\xr)}
  \notag \\
  & \psim(\xr) 
  \quad \xleftrightarrow{\text{\; forward inference \;}} \quad 
  && \pd(\xr) \; .
\label{eq:schematic}
\end{alignat}
Generative unfolding uses paired simulated events encoding a joint probability in $(\xp,\xr)$ to train a generative network with parameters $\theta$ that can sample from the needed conditional
\begin{align}
    p_\theta(\xp | \xr) \approx p(\xp|\xr) \; .
\end{align}
The unfolded distribution can then be computed via
\begin{align}
    \punf(\xp) \approx \int d\xr \, p_\theta(\xp | \xr) \pd(\xr) \; .
    \label{eq:gen_unfolding2}
\end{align}
The conditional needed for unfolding is related to the forward-simulation conditional via Bayes' theorem
\begin{align}
 p(\xp|\xr) = \frac{p(\xr|\xp) p(\xp)}{p(\xr)} \; .
\end{align}
Note that the generative model always depends on a simulation prior $p(\xp)$. In practical applications, iterative methods~\cite{Backes:2022sph, Butter:2025mek} or specific setups~\cite{Favaro:2025psi, Pazos:2024nfe} can be used to remove a potential bias arising from the simulation prior.

Current state-of-the-art generative unfolding uses conditional flow matching (CFM)~\cite{chen2018neural,lipman2023flowmatchinggenerativemodeling,albergo2023stochastic,Butter:2023fov}. It describes the transformation from a latent space $p_\text{latent}(z)$ to phase space $p(\xp)$ as a time-dependent diffusion process between $t \in [0,1]$ 
\begin{align}
    \label{eq:boundary_conditions}
    p(x,t) =  
 \begin{cases}
  p(\xp)\qqquad & t \to 0 \\
  p_\text{latent}(z) \qqquad & t \to 1  \; .
 \end{cases} 
\end{align}
The time evolution of individual phase space points is governed by a velocity field, 
\begin{align}
    \label{eq:velocity_field}
    \frac{dx(t)}{dt} = v(x(t),t)\;,
\end{align}
whereas the time evolution of the underlying probability distributions follows the continuity equation,
\begin{align}
    \frac{\partial p(x(t),t)}{\partial t} = - \nabla_x \left( v(x(t),t) p(x(t),t)\right) \;.
\end{align}
The velocity field is parametrized by a neural network,
\begin{align} 
 v_\theta (x(t),t) \approx v(x(t),t) \; .
 \label{eq:match_velocity}
\end{align}

Conditional flow matching introduces conditional target trajectories~\cite{lipman2023flowmatchinggenerativemodeling,albergo2023stochastic} that interpolate linearly between the latent space and phase space events
\begin{align}
    x(t) = tz + (1-t) \xp\;.
    \label{eq:trajectories}
\end{align}
The velocity network in Eq.\eqref{eq:match_velocity} is then trained by minimizing the mean squared error (MSE) loss 
\begin{align}
    \loss_\text{CFM}(\theta) &= \left<v_\theta(x(t),t, \xr) - v(x(t),t| z, \xp)\right>_{t\sim \mathcal{U}(0,1), \xp, \xr \sim p(\xp,\xr), z\sim p_\text{latent}(z) }^2 \notag \\ 
    &= \left<v_\theta(x(t),t, \xr) - (z-\xp) \right>_{t\sim \mathcal{U}(0,1), \xp, \xr \sim p(\xp,\xr), z\sim p_\text{latent}(z) }^2 \; ,
\end{align}
where the conditional velocity $v(x(t), t| z, \xp)$ is defined as the derivative of Eq.\eqref{eq:trajectories}.

To obtain the conditional density $p(\xp | \xr)$, paired events are sampled from the joint distribution and the detector-level event is included in the network inputs. Once training has converged, particle-level events can be sampled from the learned posterior $p_\theta(\xp|\xr)$ by numerically solving the ODE 
\begin{align}
    \xp = \int_0^1 dt \; v_\theta(x(t),t, \xr) + z
    \qquad \text{with} \quad z \sim p_\text{latent}(z), \quad \xr \sim p(\xr) \; . 
\end{align}
With the learned posterior, the unfolded distribution $\punf(\xp)$ can be obtained by (repeatedly) sampling the model given a single detector-level event, or the full unfolded distribution can be obtained by sampling the model given the full dataset.

An important strength of CFMs is that the velocity can be learned by any network architecture.
In particle physics, encoding the symmetry structure of the phase space greatly improves the network accuracy~\cite{Gong:2022lye,Bogatskiy:2022czk,Brehmer:2024yqw,Favaro:2025pgz, Witkowski:2023xmx, Nabat:2024nce,Breso:2024jlt,Bahl:2024gyt}. For this reason, the velocity is best learned by permutation equivariant transformer architectures~\cite{Heimel:2023mvw}, or transformers with an additional Lorentz-equivariant constraint~\cite{Brehmer:2024yqw,Favaro:2025pgz} like L-GATr, as illustrated in Figure~\ref{fig:money}.

In this study, both equivariance options are considered.
L-GATr includes symmetry information to enhance performance, but at the cost of slower training and additional care required in the network implementation.
Specifically, L-GATr works on geometric algebra representations and implements variants of all transformer layers that act on these representations.
The geometric algebra multivectors contain vectors and anti-symmetric tensors in addition to the scalar representations of non-equivariant transformers.
The distributions of particle-level and detector-level kinematics are not fully Lorentz-equivariant, as the symmetry is broken by various effects.
Accounting for this requires additional symmetry-breaking inputs to L-GATr.
Specifically, the beam and time directions are included as extra particles to break the Lorentz symmetry down to a residual $\text{SO}(2)$-symmetry of rotations around the beam axis.

\section{Unfolding jets and their substructure}
\label{sec:method}

Unfolding the full substructure of a jet requires predicting the kinematics of every constituent at particle level, conditioned on these quantities at detector level.
This is a set-to-set generation task in which a target set must be generated given a conditioning set, where neither set has an inherent ordering.
The multiplicity, or the number of elements in each set, need not be the same.
Unfolding jet substructure involves learning both the distribution of particle multiplicities, $\Npart$, and of particle kinematics $\xp$, conditioned on the corresponding detector-level quantities $\Ndet$ and $\xr$.
Note that the multiplicities are fully determined by the constituent kinematics through the shape of the constituent kinematics vectors, $\Npart (\xp)$ and $\Ndet (\xr)$.

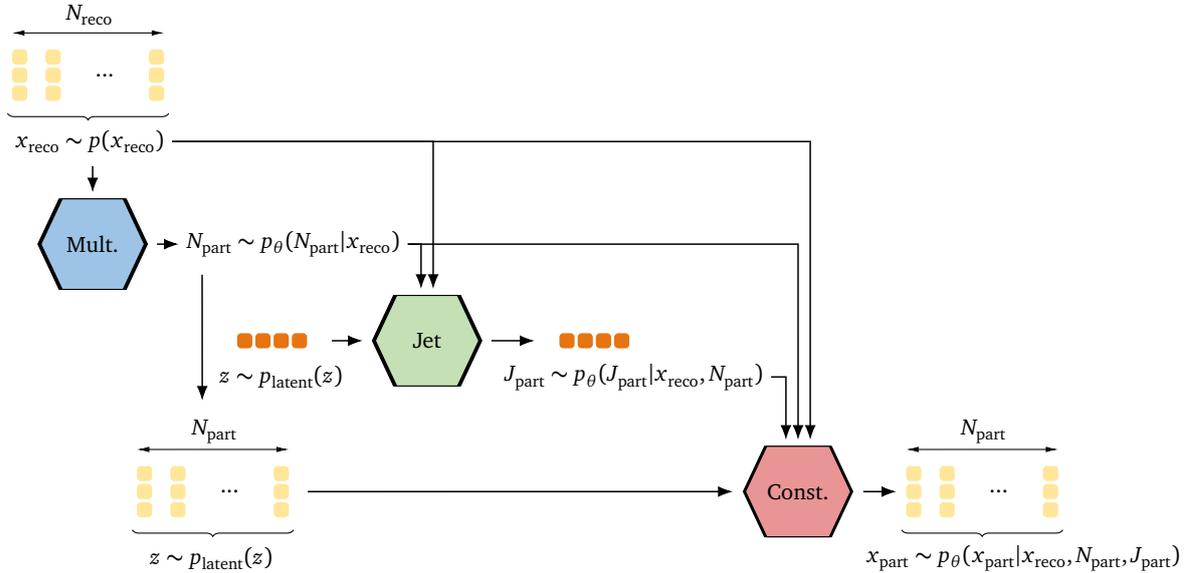
\begin{figure}[b!]
    \centering
    \usetikzlibrary{arrows, shapes.geometric, arrows.meta, shapes, decorations.pathreplacing, fit, patterns, patterns.meta,positioning,calc}

\definecolor{Rcolor}{HTML}{E99595}
\definecolor{Gcolor}{HTML}{C5E0B4}
\definecolor{Gcolor_light}{HTML}{F1F8ED}
\definecolor{Gcolor_dark}{HTML}{9CB391}
\definecolor{Bcolor}{HTML}{9DC3E6}
\definecolor{Ycolor}{HTML}{FFE699}
\definecolor{Ycolor_light}{HTML}{FFF7DE}
\definecolor{Ocolor}{HTML}{E67410}

\tikzstyle{expr} = [rectangle, rounded corners=0.3ex, minimum width=1.5cm, minimum height=1cm, text centered, align=center, inner sep=0, fill=Ycolor, font=\large, draw]
\tikzstyle{small_cinn} = [double arrow, double arrow head extend=0cm, double arrow tip angle=130, inner sep=0, align=center, minimum width=1.5cm, minimum height=1.7cm, fill=Rcolor, draw]
\tikzstyle{small_cinn_black} = [small_cinn, minimum height=1.8cm, fill=black]
\tikzstyle{transformer} = [rectangle, rounded corners, minimum width=6cm, minimum height=2.4cm, font=\large, fill=Gcolor_light, draw]
\tikzstyle{attention} = [rectangle, rounded corners=0.3ex, minimum width=5.5cm, minimum height=1.2cm, align=center, fill=Gcolor, draw, font=\large]
\tikzstyle{transformer_huge} = [rectangle, rounded corners, minimum width=1cm, minimum height=1cm, fill=Gcolor]
\tikzstyle{input} = [rectangle, rounded corners, minimum width=1cm, minimum height=1cm, draw=black]
\tikzstyle{particle_component} = [rectangle, minimum width = 0.25cm, minimum height=0.25cm, font = \large, fill=Ycolor, rounded corners=0.3ex]
\tikzstyle{attention_huge} = [rectangle, rounded corners=0.3ex, minimum width=8cm, minimum height=1.2cm, align=center, fill=Gcolor, draw, font=\large]
\tikzstyle{txt_huge} = [align=center, font=\Huge, scale=2]
\tikzstyle{txt} = [align=center, minimum height=1cm]
\tikzstyle{arrow} = [thick,-{Latex[scale=1.0]}, line width=0.2mm, color=black]
\tikzstyle{line} = [thick, line width=0.2mm, color=black]

\begin{tikzpicture}
[node distance=0.3cm, scale=0.8, every node/.style={transform shape}]
\node (particle_1_1) [particle_component]{};
\node (particle_1_2) [particle_component, below of = particle_1_1]{};
\node (particle_1_3) [particle_component, below of = particle_1_2]{};
\node (particle_2_1) [particle_component, right of = particle_1_1, xshift=0.25cm]{};
\node (particle_2_2) [particle_component, below of = particle_2_1]{};
\node (particle_2_3) [particle_component, below of = particle_2_2]{};
\node (and) [txt, right of = particle_2_2, xshift = 0.55cm]{...};
\node (particle_N_2) [particle_component, right of = and, xshift = 0.55cm]{};
\node (particle_N_1) [particle_component, above of = particle_N_2]{};
\node (particle_N_3) [particle_component, below of = particle_N_2]{};
\draw[{Latex[width=0.75mm]}-{Latex[width=0.75mm]}] ([yshift=0.4cm]particle_1_1.west) -- ([yshift=0.4cm]particle_N_1.east);
\draw [decorate,decoration={brace, mirror}]
  ([yshift=-0.2cm, xshift=-0.2cm]particle_1_3.south) -- ([yshift=-0.2cm, xshift=0.2cm]particle_N_3.south){};
\node (N_reco) [txt, above of = and, yshift=0.7cm, xshift=-0.25cm]{$\Ndet$};
\node (xreco)[txt, below of = N_reco, yshift=-1.8cm]{$\xr \sim p(\xr)$};

\node (Multi_b)[small_cinn_black, below of = and, xshift=-0.2cm, yshift=-2.5cm]{};
\node (Multi)[small_cinn, above of=Multi_b, yshift=-0.3cm, fill=Bcolor]{Mult.};
\node (xmult) [txt, right of=Multi, xshift=3.0cm]{$\Npart \sim p_\theta (\Npart|\xr)$};

\node (CFM_b_jet) [small_cinn_black, right of = xmult, xshift=1.9cm, yshift=-1.6cm]{};
\node (CFM_jet) at (CFM_b_jet) [small_cinn, fill=Gcolor]{Jet};

\node (base_jet_1) [particle_component,left of = CFM_jet, xshift=-1.8cm, fill=Ocolor]{};
\node (base_jet_2) [particle_component,left of = base_jet_1, fill=Ocolor]{};
\node (base_jet_3) [particle_component,left of = base_jet_2, fill=Ocolor]{};
\node (base_jet_4) [particle_component,left of = base_jet_3, fill=Ocolor]{};
\node (xbasejet)[txt, below of=base_jet_1, yshift=-0.3cm, xshift=-0.3cm]{$z \sim p_{\text{latent}}(z)$};

\node (final_jet_1) [particle_component,right of = CFM_jet, xshift=2.0cm, fill=Ocolor]{};
\node (final_jet_2) [particle_component,right of = final_jet_1, fill=Ocolor]{};
\node (final_jet_3) [particle_component,right of = final_jet_2, fill=Ocolor]{};
\node (final_jet_4) [particle_component,right of = final_jet_3, fill=Ocolor]{};
\node (xjet)[txt, below of = final_jet_1, xshift=1.1cm, yshift=-0.3cm]{$\Jpart \sim p_\theta(\Jpart|\xr, \Npart)$};

\node (CFM_b) [small_cinn_black, right of = CFM_jet, xshift=5.8cm, yshift=-2.5cm]{};
\node (CFM) at (CFM_b) [small_cinn]{Const.};

\node (base_node) [txt, left of=CFM, xshift=-9.05cm]{...};
\node (base_2_2) [particle_component,left of = base_node, xshift=-0.55cm, fill=Ycolor]{};
\node (base_2_1) [particle_component,above of = base_2_2, fill=Ycolor]{};
\node (base_2_3) [particle_component,below of = base_2_2, fill=Ycolor]{};
\node (base_1_2) [particle_component,left of = base_2_2, xshift=-0.25cm, fill=Ycolor]{};
\node (base_1_1) [particle_component, above of = base_1_2 , fill=Ycolor]{};
\node (base_1_3) [particle_component, below of = base_1_2, fill=Ycolor]{};
\node (base_N_2) [particle_component,right of = base_node,xshift = 0.55cm, fill=Ycolor]{};
\node (base_N_1) [particle_component,above of = base_N_2, fill=Ycolor]{};
\node (base_N_3) [particle_component,below of = base_N_2, fill=Ycolor]{};
\node (pgauss) [txt, below of = base_node, yshift=-0.8cm, xshift=-0.3cm]{$z\sim p_\text{latent}(z)$};
\node (N_reco) [txt, above of = base_node, yshift=0.7cm, xshift=-0.25cm]{$\Npart$};
\draw [decorate,decoration={brace, mirror}]
  ([yshift=-0.2cm, xshift=-0.2cm]base_1_3.south) -- ([yshift=-0.2cm, xshift=0.2cm]base_N_3.south){};
\draw[{Latex[width=0.75mm]}-{Latex[width=0.75mm]}] ([yshift=0.4cm]base_1_1.west) -- ([yshift=0.4cm]base_N_1.east);

\node (final_node) [txt, right of=CFM, xshift=3cm]{...};
\node (final_2_2) [particle_component,left of = final_node, xshift=-0.55cm, fill=Ycolor]{};
\node (final_2_1) [particle_component,above of = final_2_2, fill=Ycolor]{};
\node (final_2_3) [particle_component,below of = final_2_2, fill=Ycolor]{};
\node (final_1_2) [particle_component,left of = final_2_2, xshift=-0.25cm, fill=Ycolor]{};
\node (final_1_1) [particle_component, above of = final_1_2 , fill=Ycolor]{};
\node (final_1_3) [particle_component, below of = final_1_2, fill=Ycolor]{};
\node (final_N_2) [particle_component,right of = final_node,xshift = 0.55cm, fill=Ycolor]{};
\node (final_N_1) [particle_component,above of = final_N_2, fill=Ycolor]{};
\node (final_N_3) [particle_component,below of = final_N_2, fill=Ycolor]{};
\node (ppart) [txt, below of = final_node, yshift=-0.8cm, xshift=0.4cm]{$\xp \sim p_\theta(\xp|\xr, \Npart, \Jpart)$};
\node (N_reco2) [txt, above of = final_node, yshift=0.7cm, xshift=-0.25cm]{$\Npart$};
\draw [decorate,decoration={brace, mirror}]
  ([yshift=-0.2cm, xshift=-0.2cm]final_1_3.south) -- ([yshift=-0.2cm, xshift=0.2cm]final_N_3.south){};
\draw[{Latex[width=0.75mm]}-{Latex[width=0.75mm]}] ([yshift=0.4cm]final_1_1.west) -- ([yshift=0.4cm]final_N_1.east);

\draw [arrow, color=black] ([yshift=0.1cm,xshift=0.05cm]xreco.south) -- ([yshift=0.1cm]Multi.north);
\draw [arrow, color=black] ([xshift=0.2cm]Multi.east) -- (xmult.west);

\draw [arrow, color=black] (xreco.east) -- ([xshift=0.1cm]CFM_jet |- xreco.east) -- ([yshift=0.1cm, xshift=0.1cm]CFM_jet.north);
\draw [arrow, color=black] (xmult.east) -- ([xshift=-0.1cm]CFM_jet |- xmult.east) -- ([yshift=0.1cm, xshift=-0.1cm]CFM_jet.north);
\draw [arrow, color=black] ([xshift=0.4cm]base_jet_1.east) -- ([xshift=-0.2cm]CFM_jet.west);
\draw [arrow, color=black] ([xshift=0.2cm]CFM_jet.east) -- ([xshift=-0.4cm]final_jet_1.west);

\draw [arrow, color=black] (xreco.east) -- ([xshift=0.2cm]CFM |- xreco.east) -- ([yshift=0.1cm, xshift=0.2cm]CFM.north);
\draw [arrow, color=black] (xmult.east) -- (CFM |- xmult.east) -- ([yshift=0.1cm]CFM.north);
\draw [arrow, color=black] (xjet.east) -- ([xshift=-0.2cm]CFM |- xjet.east) -- ([xshift=-0.2cm,yshift=0.1cm]CFM.north);

\draw [arrow, color=black] ([xshift=1.0cm]base_node.east) -- ([xshift=-0.2cm]CFM.west);
\draw [arrow, color=black] ([xshift=0.2cm]CFM.east) -- ([xshift=-1.4cm]final_node.west);
\draw [arrow, color=black] ([xshift=-1.5cm]xmult.south) -- ([xshift=-0.2cm]N_reco.north);

\end{tikzpicture}
    \caption{Three-step generative substructure unfolding: multiplicity, jet, and jet constituents.}
    \label{fig:model}
\end{figure}

The high-dimensional unfolding task is divided into three steps, as illustrated in Figure~\ref{fig:model}.
\begin{enumerate}
\item First the multiplicity $\Npart$ is unfolded with a parametrized ansatz for its distribution;
\item Second,  the jet kinematics $\Jpart$ are predicted using a CFM network;
\item Finally, the complete constituent kinematics $\xp$ are unfolded, again using CFM.
\end{enumerate}
This corresponds to an autoregressive decomposition of the jet substructure probability distribution
\begin{align}
    p_\theta(\xp, \Jpart, \Npart | \xr) =& p_\theta(\Npart | \xr) \notag\\
    &\times p_\theta(\Jpart | \xr, \Npart) \notag \\
    &\times p_\theta(\xp | \xr, \Npart, \Jpart)\;.
\end{align}
For clarity, the unfolded multiplicity $\Npart$ and jet kinematics $\Jpart$ are included on the left-hand side, although both are implicitly determined by $\xp$. The corresponding log-likelihood loss reads
\begin{align}
    \loss_\text{unfold}
    =& - \left\langle \log p_\theta(\xp, \Jpart, \Npart | \xr)\right\rangle_{\xr \sim p(\xr), \xp\sim p(\xp | \xr)} \notag \\
    =& - \left\langle \log p_\theta(\Npart | \xr )\right\rangle_{\xr\sim p(\xr), \xp\sim p(\xp | \xr)} \notag\\
    & - \left\langle \log p_\theta(\Jpart | \xr, \Npart)\right\rangle_{\xr\sim p(\xr), \xp\sim p(\xp | \xr)} \notag\\
    & - \left\langle \log p_\theta(\xp | \xr, \Npart, \Jpart)\right\rangle_{\xr\sim p(\xr), \xp\sim p(\xp|\xr)} .
\label{eq:loss_comb}
\end{align}
This blueprint for the loss function allows an independent training of the three unfolding networks.
However, the generative networks for $\Jpart$ and $\xp$ are trained using CFM instead of a log-likelihood loss.
The three networks are chained together at inference time to sample $\xp$, see Figure~\ref{fig:model}. The three unfolding steps are described in Sections~\ref{sec:multiplicity}, \ref{sec:jet-unfolding}, and~\ref{sec:constituent-unfolding}. The benefit of this factorization into learned conditionals is that this new method can be optimized to accurately reproduce jet-level observables and substructure observables separately, such that the jet-level unfolding does not suffer from the substructure curse of dimensionality.

\subsection{Multiplicity unfolding}
\label{sec:multiplicity}

Experiments at the LHC do not reconstruct every particle that is produced in a collision.
Particles can fall outside of the detector acceptance or otherwise fail to be reconstructed, or fake particles can be reconstructed due to random alignments of detector hits. Hence,  the number of detector-level $\Ndet$ and particle-level $\Npart$ constituents are not identical. As detector effects are stochastic, the multiplicity at particle-level is modeled as the conditional distribution
\begin{align}
    \label{eq:conditional_particle_mulitplicity}
    p(\Npart | \xr) \;,
\end{align}
which we approximate using a mixture of (five) Gaussian distributions with learnable means $\mu_{\theta,i}(\xr)$, widths $\sigma_{\theta,i}(\xr)$ and weights $w_{\theta, i}(\xr)$, such that 
\begin{align}
    \label{eq:learned_multiplicity}
    p_\theta(\Npart| \xr) =  \sum_{i=1}^5 w_{\theta,i} \mathcal{N}(\mu_{\theta ,i}, \sigma_{\theta,i})
    \qquad \text{with} \quad \sum_{i=1}^5 w_{\theta,i} = 1 \;.
\end{align}
The condition on $w_{\theta, i}$ ensures normalization. The explicit $\xr$ dependence of the Gaussian coefficients is omitted for readability. The coefficients are predicted by a neural network with parameters $\theta$, which following Eq.\eqref{eq:loss_comb} are optimized by a log-likelihood loss
\begin{align}
    \loss_{\text{Multiplicity}} = \langle -\log p_\theta(\Npart|\xr) \rangle_{ \xr \sim p(\xr), \Npart \sim p(\Npart | \xr)} \; .
\end{align}
For a given detector-level event $\xr$, the multiplicity predictor can be evaluated by sampling from the learned distribution in Eq.\eqref{eq:learned_multiplicity} and rounding the sampled value to the nearest integer, to predict $\Npart$.

For the multiplicity unfolding, a transformer encoder is used. The input tokens are the detector-level constituents $\xr$, each represented by their four-momenta $(E, p_x,p_y, p_z)$. 
A stack of transformer blocks processes the inputs into a 15-dimensional space, then each dimension is averaged over all tokens to extract 15 numbers.
These are interpreted as the parameters of the Gaussian mixture model in Eq.\eqref{eq:learned_multiplicity}, after applying an exponential function on all $\sigma_{\theta,i}$ and a softmax function on the $w_{\theta,i}$.

For the L-GATr architecture, the constituent kinematics are embedded into the vector parts of geometric algebra representations. The Gaussian mixture model parameters are extracted from the scalar parts of the geometric algebra representations. The beam and time directions are embedded into multivectors and added to the token list to allow for symmetry breaking. This encoder architecture is used with slight modifications to process the detector-level conditions for the jet unfolding and constituent unfolding networks.

\subsection{Jet unfolding}
\label{sec:jet-unfolding}

The jet kinematics are unfolded and used to pre-process the constituent kinematics $\xp$ such that they are generated relative to the jet $\Jpart$. This method could be extended to unfold entire events by replacing the jet kinematics with some common reference, for example the vector boson in $V$+jets events. While it is possible to unfold the constituent kinematics directly as $p(\xp | \xr, \Npart)$, unfolding the jet kinematics first helps the generative model that performs the constituent unfolding capture challenging substructure correlations with high accuracy. The particle-level jet kinematics are parametrized as
\begin{align}
  \Jpart = (\log p_{T,J}, \phi_J, \eta_J, \log m_J^2) \; ,
  \label{eq_jpart}
\end{align}
with each component standardized. The network receives the multiplicity $\Npart$ as well as the detector-level jet kinematics $\Jdet$ as conditions. Though the detector level jet kinematics can be sensitive to ``hidden-variables'' which produce a non-universal forward simulation~\cite{Butter:2025via}, in this study it is sufficient to use only these inputs instead of the full detector level constituent kinematics information $\xr$. Additionally, alternative parametrizations of the jet kinematics $\Jpart$ can help the network to learn task-specific features. See Section~\ref{sec:z_plus_jets}.

For the jet kinematics, a transformer encoder processes the detector-level condition. $\Jpart$ is embedded as in Eq.\eqref{eq_jpart}, with each component standardized, together with the detector-level and particle-level multiplicities $\Ndet$ and $\Npart$ embedded using Gaussian Fourier projections.

A transformer decoder then predicts the velocity field
\begin{align}
    v_\theta(\Jpart (t), t, \Npart, \Jdet)\;.
    \label{eq:jet_kinematics_velocity}
\end{align}
The latent CFM distribution is chosen as uniform for $\phi_J$ and unit Gaussians for $\log p_{T,J}$, $\eta_J$, $\log m_J^2$. The decoder operates on four tokens, one for each component of the particle-level jet, which are distinguished from each other using a one-hot encoding. The time $t$ is embedded using a Gaussian Fourier projection. The detector-level jet information is included in each transformer decoder block through cross-attention. Four velocity components are extracted from the four tokens and identified as $(v_{\log p_{T,J}}, v_{\phi_J}, v_{\eta_J}, v_{\log m_J^2})$.

For the L-GATr, the condition is embedded using an encoder similar to the multiplicity unfolding, but now including only the four-momentum of the detector-level jet $\Jdet$ and constituent multiplicity $\Ndet$ instead of the full constituent information $\xr$. The beam and time directions are also included to allow for symmetry breaking. The decoder for the velocity prediction takes the noised particle-level jet four-momentum, parametrized in rectilinear coordinates as $(E_J, p_{x,J}, p_{y,J}, p_{z,J})$ as input, together with the time embedded in a scalar channel, and returns a four-velocity $(v_{E_J}, v_{p_{x,J}}, v_{p_{y,J}}, v_{p_{z,J}})$. The linear CFM trajectories are defined in the $(\log p_{T,J}, \phi_J, \eta_J, \log m_J^2)$ parametrization, so the particle-level jet $\Jpart$ is transformed to rectilinear coordinates before input to the flow matching network. After extracting the four-velocity and multiplying by  the inverse Jacobian of the forward transformation, the velocity field $(v_{\log p_{T,J}}, v_{\phi_J} v_{\eta_J}, v_{\log m_J^2})$ is obtained. To avoid numerical instabilities from large Jacobians, after the transformation, the $v_{\log p_{T,J}}$ and $v_{\log m_J^2}$ components are over-written with additional Lorentz scalar outputs of the L-GATr network, while the $v_{\phi_J}$ and $v_{\eta_J}$ components are kept~\cite{Brehmer:2024yqw}.

\subsection{Constituent unfolding}
\label{sec:constituent-unfolding}

\begin{figure}[t]
    \includegraphics[width=0.325\textwidth,page=47]{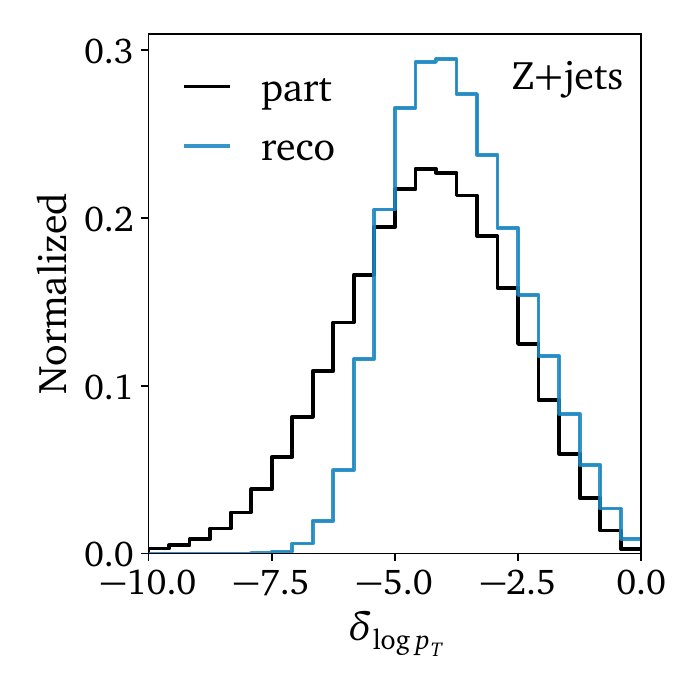}
    \includegraphics[width=0.325\textwidth,page=48]{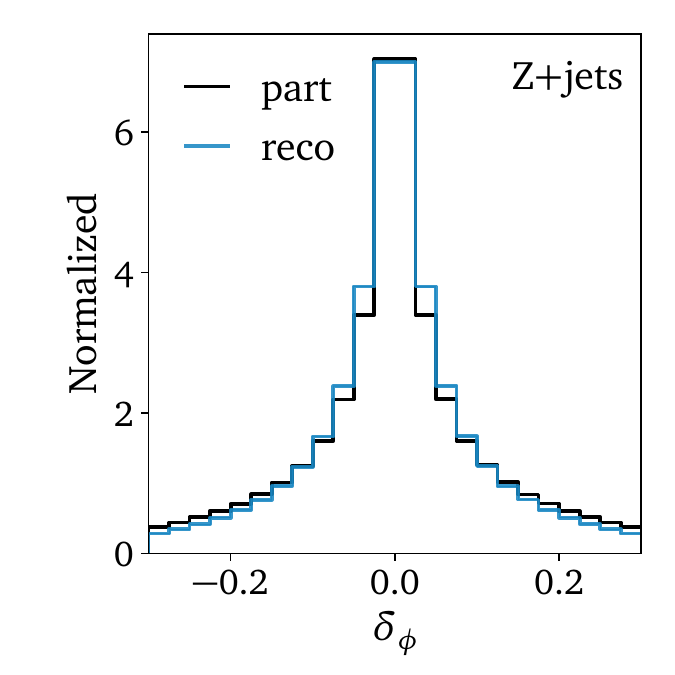}
    \includegraphics[width=0.325\textwidth,page=49]{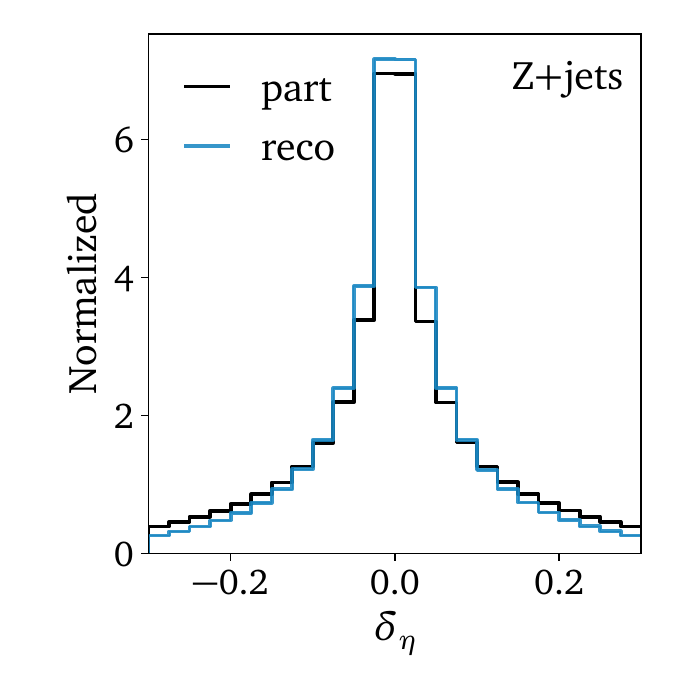}
    \caption{Constituents distributions for Z+jets events in the $(\delta_{\log p_T}, \delta_\phi, \delta_\eta)$ parametrization at particle level and detector level.}
    \label{fig:z_dataset_distributions}
\end{figure}

The predicted particle-level jet kinematics enable a physically motivated pre-processing to be applied that represents the particle-level constituent kinematics relative to the particle-level jet $\Jpart$,
%
\begin{align}
    (\delta_{\log p_T}, \delta_\phi, \delta_\eta ) = (\log p_T - \log p_{T,J}, \phi - \phi_J, \eta - \eta_J )\;.
    \label{eq:relative-parametrization}
\end{align}
As shown in Figure~\ref{fig:z_dataset_distributions} for boosted $Z$ plus jet production, the distributions of $\delta_\phi$ and $\delta_\eta$ are sharply peaked around zero, since most constituents are aligned with the jet-axis. Using this parametrization in the constituent unfolding task is found to significantly improve the accuracy of the unfolded jet substructure. The particle-level jet kinematics are used to produce this parametrization and are given as an additional condition to sample the particle-level density. However, the unfolded constituent kinematics fully capture the particle-level density. The jet kinematics are only a requirement to produce them and are discarded afterwards.

The conditional network framework developed in the last section is adapted to also generate the particle-level constituents. In this case the phase space corresponds to the particle-level constituent kinematics $\xp$ from Eq.\eqref{eq:relative-parametrization}. The distributions of each component of these 3-vectors are preprocessed to have zero mean and unit variance before training. The conditions are the particle-level $\Npart$ and $\Jpart$, as well as the detector-level information $\xr$. The velocity field now takes the form
\begin{align}
    \label{eq:v_constituents}
    v_\theta(\xp (t), t, \Npart, \Jpart, \xr )\;.
\end{align}
The latent distribution $p_\text{latent}(z)$ is the product of three Gaussian distributions for $\delta_{\log p_T}$, $\delta_\phi$, $\delta_\eta$, where the Gaussian for $\delta_\phi$ is clamped to take into account the $2\pi$-periodicity.

The constituent encoder-decoder structure follows the jet unfolding network, but with a large number of tokens at the encoder and decoder level. The full detector-level condition $\xr$ is processed with a transformer encoder, now including the full detector-level constituent information for both transformer architectures. The detector-level jet kinematics $\Jdet$ are included as an additional token embedded in the form $(p_{T,J}, \phi_J, \eta_J, m_J^2)$ and distinguished from the constituents through a one-hot encoding. The encoder now operates on fixed-mass jet constituents instead of the jet whose mass can shift between particle and detector level. The constituent masses are fixed, and the corresponding velocities are not learned. Many particle-level constituents are generated instead of a single jet, with one token per particle in the transformer decoder. The particle-level jet kinematics $\Jpart$ are also included as an additional token in the decoder and embedded in the same way as the constituent kinematics, but again distinguished through a one-hot encoding. In contrast to the constituent tokens in the decoder, this jet token is not time dependent and is masked out when extracting the velocity in Eq.\eqref{eq:v_constituents}. In addition to the constituent representation discussed in Section~\ref{sec:constituent-unfolding}, a sinusoidal encoding of the $p_T$-ordering of the jet constituents is included, breaking permutation equivariance.

The L-GATr architecture for constituent-level unfolding extends the constituent unfolding transformer by the same steps as the jet-level L-GATr extends the jet-level transformer described above.

\section{Unfolding jets produced with a Z boson}
\label{sec:z_plus_jets}

The first task considered is to unfold the substructure of jets in a benchmark unfolding dataset, containing jets taken from a simulation of the process
\begin{align}
  pp \to Z + \text{jets}
  \qquad \text{with} \quad p_{\text{T}}^Z > 150 \,\text{GeV} \; ,
\end{align}
both at detector- and particle-level, with $\sqrt{S} = 14 \;\text{TeV}$~\cite{Andreassen:2019cjw}. The hard process, parton showering and hadronization are simulated using \pythia 8.243~\cite{Sjostrand:2014zea}, and the detector response in \delphes 3.4.2~\cite{deFavereau:2013fsa} with the default CMS detector card. \delphes provides reconstructed particle flow objects which can be assigned to detector-level particles.
Detector- and particle-level constituents are clustered into jets using the anti-k$_\text{T}$ algorithm~\cite{Cacciari:2008gp} with $R=0.4$ in \fastjet~\cite{Cacciari:2011ma}.
The dataset contains 1.6 million events, of which 80\% are used for training.
The performance of the unfolding is evaluated on the remaining 20\%. 

\begin{figure}[b!]
    \includegraphics[width=0.325\textwidth,page=1]{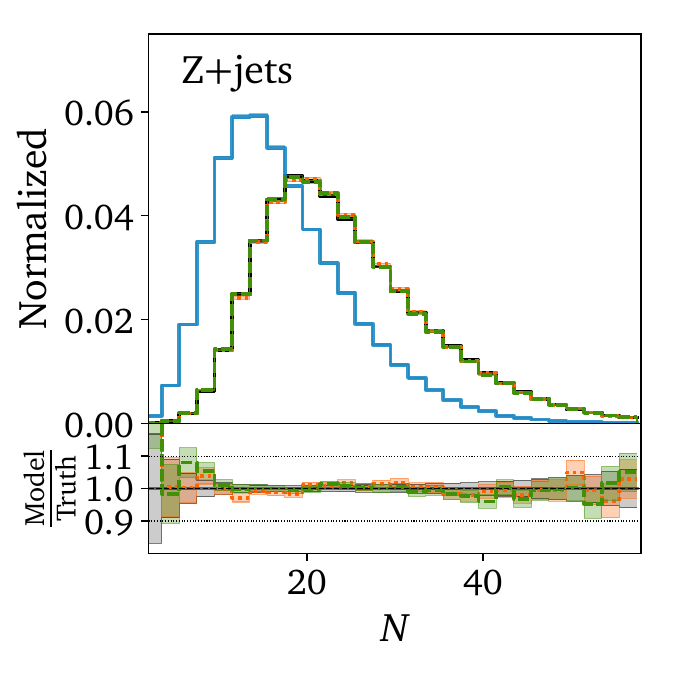}
    \includegraphics[width=0.325\textwidth,page=2]{plots/plots_z.pdf}
    \includegraphics[width=0.325\textwidth,page=5]{plots/plots_z.pdf}
    \caption{Unfolded distributions for the jet multiplicity as predicted by the multiplicity unfolding network and the jet transverse momentum and mass predicted directly with the jet unfolding network.}
    \label{fig:z_mult_jet}
\end{figure}

Following Section~\ref{sec:method}, the multiplicity predictor network is trained first.  In the left panel of Figure~\ref{fig:z_mult_jet}, a clear shift is observed between particle-level and detector-level multiplicity, with the particle-level multiplicity having a substantially higher mean. Both the transformer and L-GATr based multiplicity predictor networks reliably predict the particle-level multiplicity and match the true conditional density to percent-level accuracy.

Next, the predicted densities of the jet $p_T$ and mass provided by the jet unfolding model are shown in the central and right panels of Figure~\ref{fig:z_mult_jet}.  For this unfolding task in particular, the jet transverse momentum is parametrized as
\begin{align}
    p_{T,J} \quad \to\quad \text{arcsinh}\frac{p_{T,J} - 200\;\text{GeV}}{10\;\text{GeV}}\;
\end{align}
which was found to produce better performance than the logarithm based parametrization shown in Section~\ref{sec:jet-unfolding}. The predicted densities show excellent agreement with the true particle-level densities. These quantities are only used to perform the pre-processing in Eq.\eqref{eq:relative-parametrization}, which expresses the momenta of the constituents relative to the particle-level jet.

\begin{figure}[t]
    \includegraphics[width=0.325\textwidth,page=6]{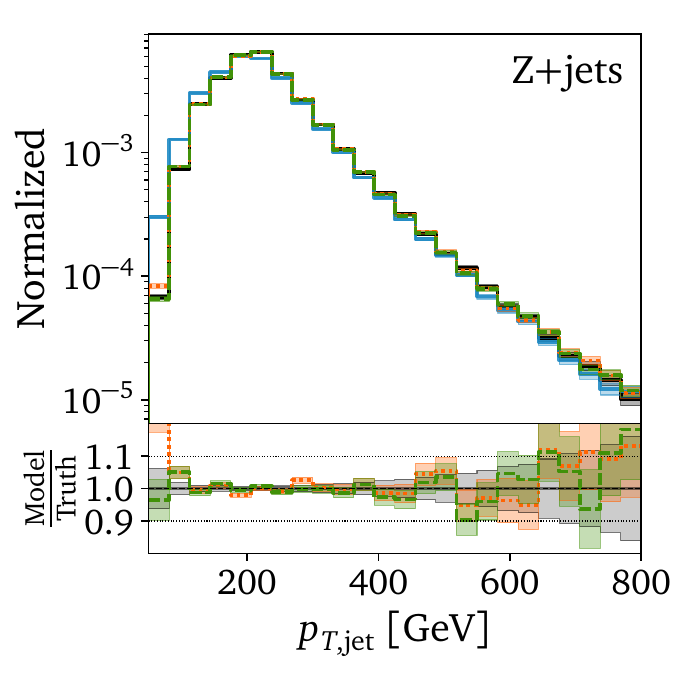}
    \includegraphics[width=0.325\textwidth,page=8]{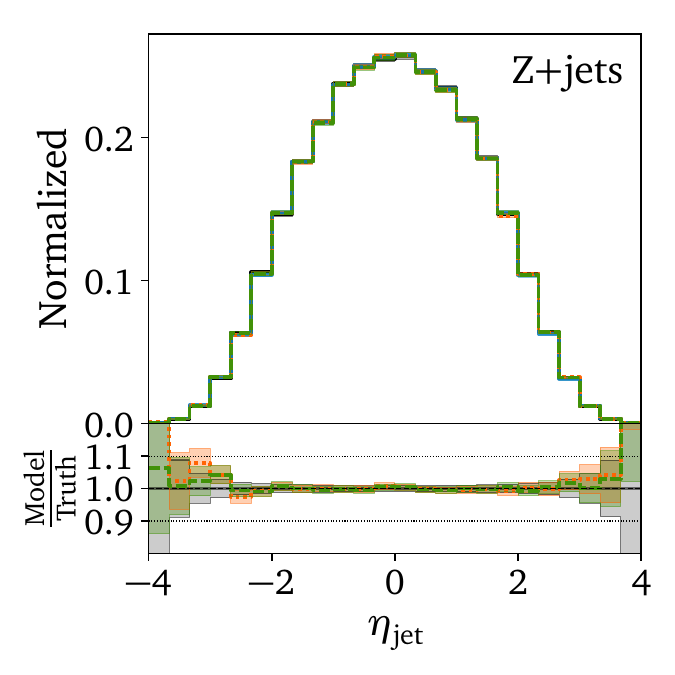}
    \includegraphics[width=0.325\textwidth,page=9]{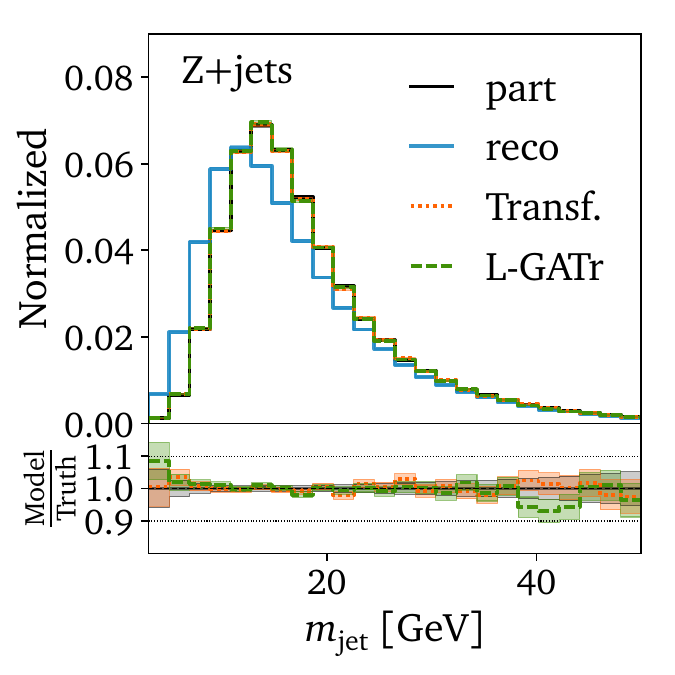}\\
    \includegraphics[width=0.325\textwidth,page=30]{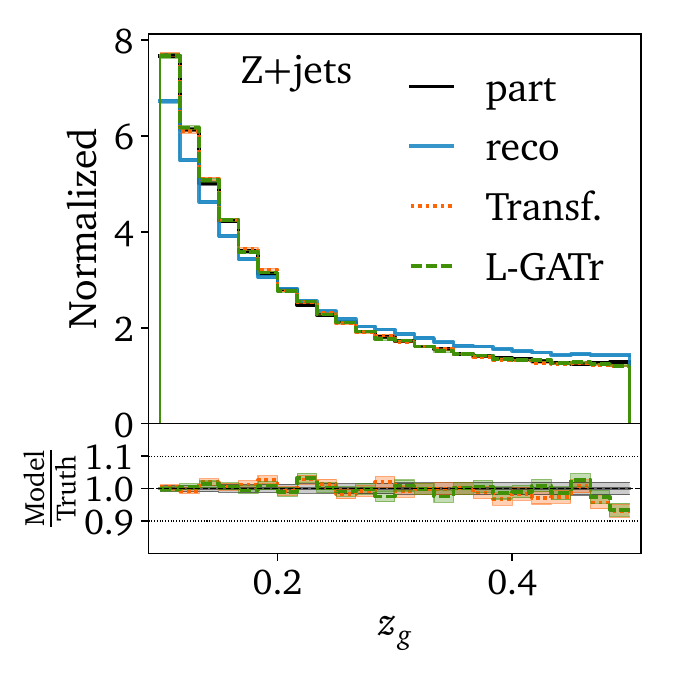} 
    \includegraphics[width=0.325\textwidth,page=29]{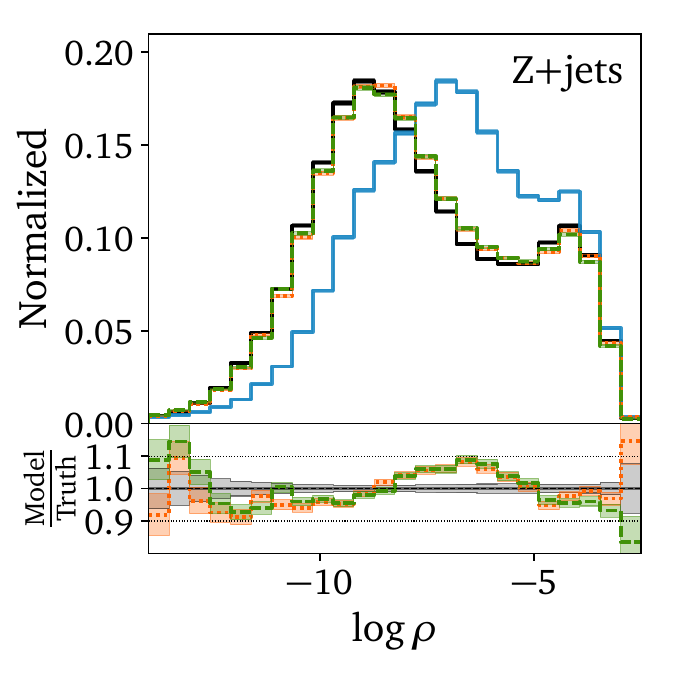}
    \includegraphics[width=0.325\textwidth,page=46]{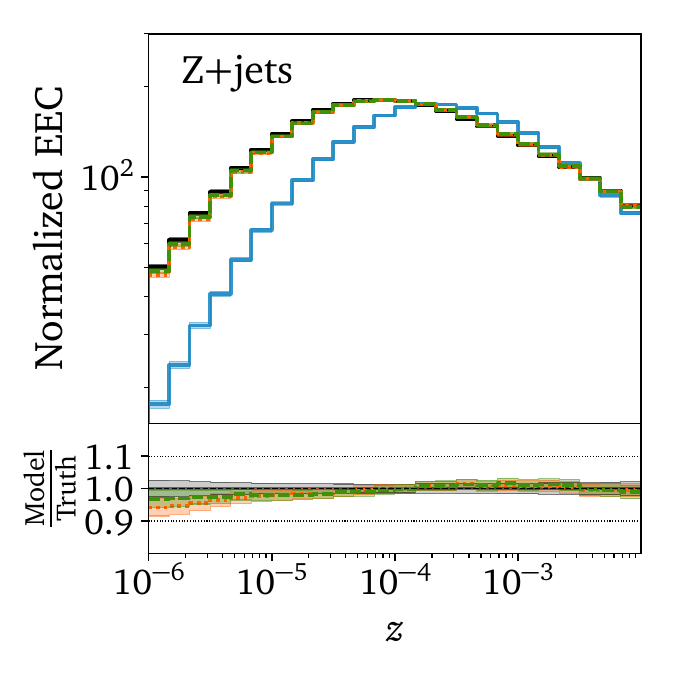}
    \caption{Distributions of jet (substructure) observables computed from the unfolded constituents.}
    \label{fig:z_constituents1}
\end{figure}

Finally, all jet constituents are unfolded.  The top panels of Figure~\ref{fig:z_constituents1} show jet-level observables computed from the unfolded constituent 4-momenta. Although the jet 4-momentum is now a high-dimensional correlation between all constituents, the jet $p_T$, mass, and pseudo-rapidity are unfolded to percent level accuracy. This accuracy is enabled by the pre-processing provided by the jet unfolding. Training the flow matching network with a naive parametrization of the constituent kinematics is found to produce significant deviations from the truth distributions.

\begin{figure}[t]
    \includegraphics[width=0.325\textwidth,page=22]{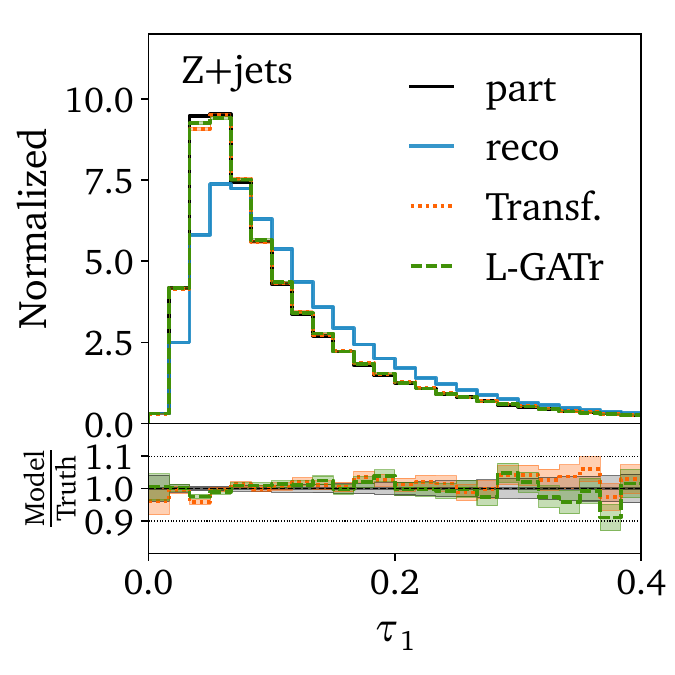}
    \includegraphics[width=0.325\textwidth,page=23]{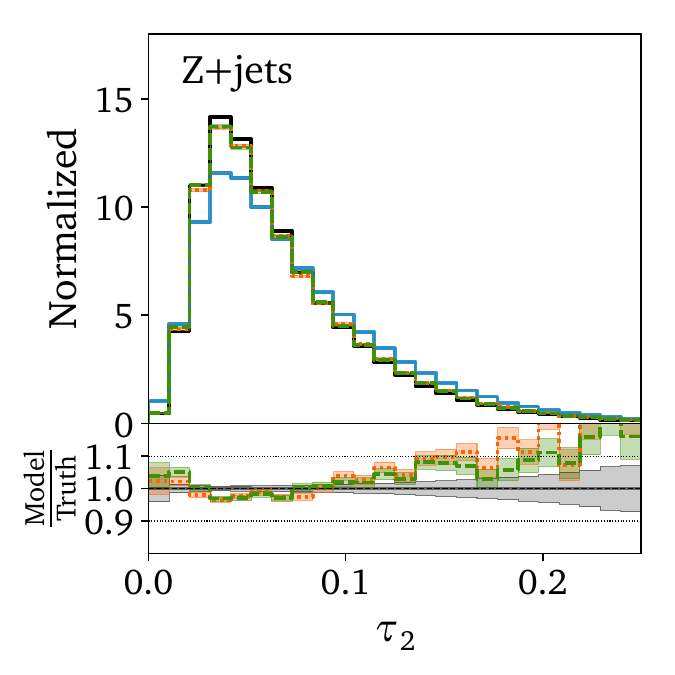}
    \includegraphics[width=0.325\textwidth,page=26]{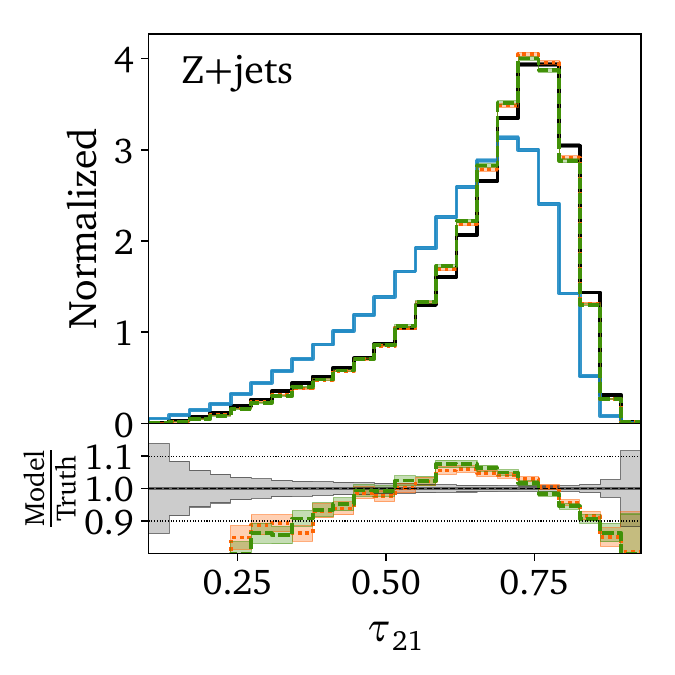}
    \caption{Challenging distributions of jet observables computed from the unfolded constituents.}
    \label{fig:z_constituents2}
\end{figure}

To examine the accuracy of the unfolding of the entire substructure of the jet, the generated and true particle-level densities for several substructure observables are shown in the lower panels of Figure~\ref{fig:z_constituents1} and in Figure~\ref{fig:z_constituents2}. Accurate unfolding of these observables with a generative model was previously only possible through unfolding them directly, as opposed to first unfolding the full substructure of the jet and calculating them post hoc.

The selection of observables~\cite{Andreassen:2019cjw} in Figure~\ref{fig:z_constituents1} includes the groomed momentum fraction $z_g$~\cite{Larkoski:2014wba} the logarithm of soft drop mass $\log \rho$, and the two-point energy correlator (EEC) over the distance between constituents $z$~\cite{Moult:2025nhu} in the second row. The top panels confirm that the jet-level observables from Figure~\ref{fig:z_mult_jet} are reproduced accurately in terms of the constituents. The momentum fraction and and EEC show percent level agreement between the unfolded and true particle-level distributions, deviations are limited to tails with limited statistics. For the soft drop mass the peak position is very slightly shifted, turning into vertical deviations in the binned distribution of up to 10\% on the shoulders of the peak. For all distributions in Figure~\ref{fig:z_constituents1} we observe no difference in performance between L-GATr and the Transformer, indicating that all architectures work well with the given hyperparameters.

Figure~\ref{fig:z_constituents2} shows the jet width $w \equiv \tau_1^{\beta=1}$, the 2-subjettiness $\tau_2^{\beta=1}$, and the 21-subjettiness ratio $\tau_{21}^{\beta=1}$~\cite{Thaler:2010tr}. While the jet width is reproduced at the percent level again, the 2-subjettiness $\tau_2$ as a proper angular correlation proves to be more challenging, and our deep-routed scientific honesty motivates us to also show the one observable which remains a challenge for both transformer architectures and the current network training. This is $\tau_{21}$, with deviations in the 5\% to 10\% even for the  bulk shape. 

\section{Unfolding boosted top jets}
\label{sec:ttbar}

The precision-generative unfolding method is additionally applied to the task of unfolding the substructure of jets originating from the all-hadronic decay of boosted top quarks. Boosted hadronically decaying top jets have become a standard analysis object at the LHC~\cite{Plehn:2009rk,Plehn:2011tg} and can even be used to extract the top mass~\cite{CMS:2022kqg,Holguin:2024tkz,Favaro:2025psi}.

The increased constituent multiplicity and more-complicated substructure arising from the three-body decay of the top quark make top jets a more challenging unfolding task. The hard process
\begin{align}
    p p \rightarrow t \bar{t} \rightarrow (b q\bar{q}^\prime) \,  (\bar{b} \ell \bar{\nu}) + \text{c.c},
\end{align}
is generated with \madgraph 5~\cite{Alwall:2011uj} assuming a top mass of $m_t = 173$ \gev. Parton showering and hadronization are simulated with \pythia  8.313~\cite{Sjostrand:2014zea} and the detector response is simulated with \delphes 3.5.0~\cite{deFavereau:2013fsa} using the default CMS card.
After reconstruction, detector-level and particle-level jets are defined via the anti-$k_\text{T}$ algorithm~\cite{Cacciari:2008gp} with $R=1.2$ using \fastjet~\cite{Cacciari:2011ma}. Each event is required to have exactly two jets with $p_{\text{T}}^J > 10$ GeV and exactly one lepton satisfying $p_{\text{T}}^\ell > 60$ GeV and $|\eta^{\ell}| < 2.4$.
In addition, each event must have exactly two $b$-tagged sub-jets, clustered with anti-$k_\text{T}$ and $R=0.4$. The jet with the largest angular distance to the lepton is tagged as the top jet. The top jet is additionally required to satisfy
\begin{align}
  p_{\text{T}}^J > 400~\text{GeV}
  \qquad \text{and} \qquad
  |\eta^J| < 2.4 \; ,
\end{align}
and to have exactly one of the two $b$-tagged subjets within the top jet. Except for the cuts based on the $b$-tagging information, which is only available at detector level, all cuts are applied to both detector level and particle level. After applying all cuts, there are 6M top jets in the dataset. 90\% of the events are used for training and the remaining 10\% for evaluation.

\begin{figure}[t]
    \includegraphics[width=0.325\textwidth,page=2]{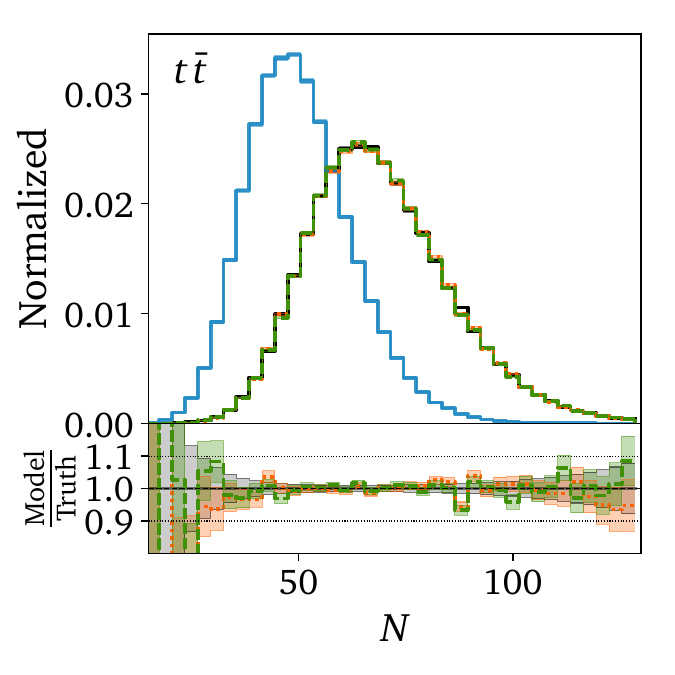}
    \includegraphics[width=0.325\textwidth,page=7]{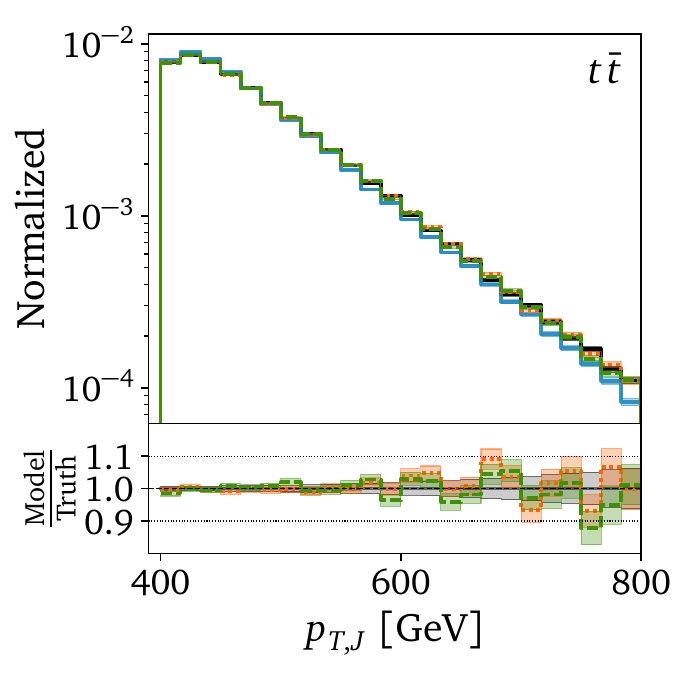}
    \includegraphics[width=0.325\textwidth,page=10]{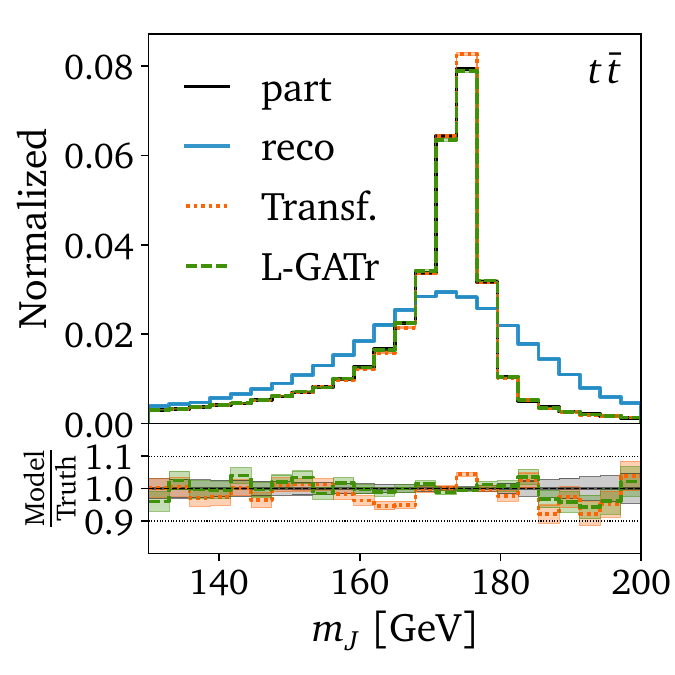}
    \caption{Unfolded distributions for the jet multiplicity as predicted by the multiplicity unfolding network and the jet transverse momentum and mass predicted directly with the jet unfolding network.}
    \label{fig:t_mult_jet}
\end{figure}

\begin{figure}[t!]
    \includegraphics[width=0.325\textwidth,page=11]{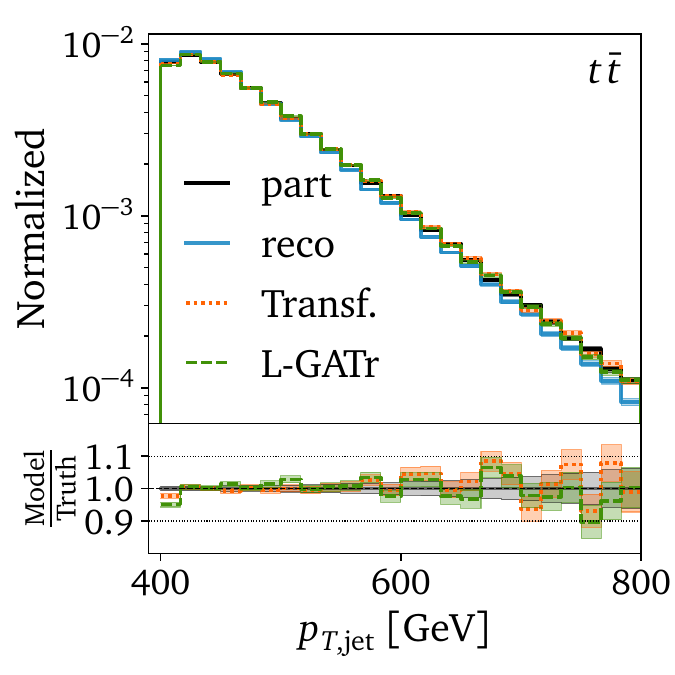}
    \includegraphics[width=0.325\textwidth,page=13]{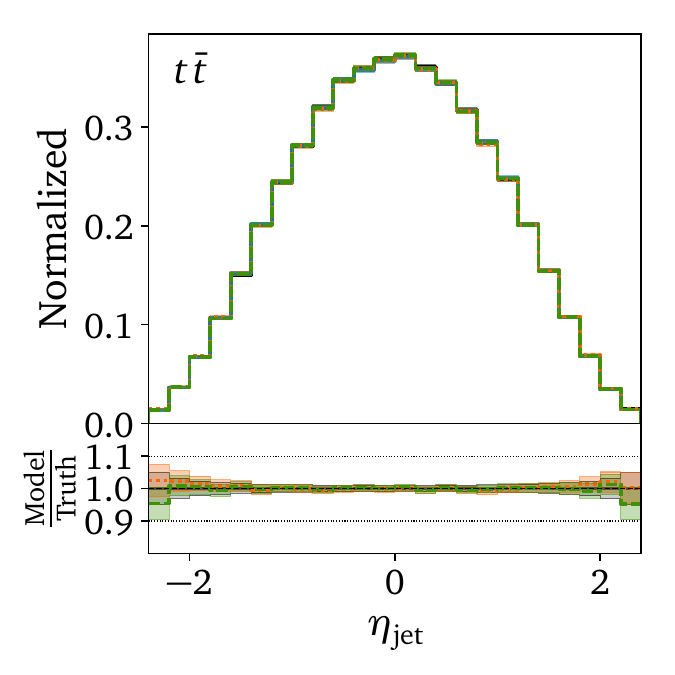}
    \includegraphics[width=0.325\textwidth,page=14]{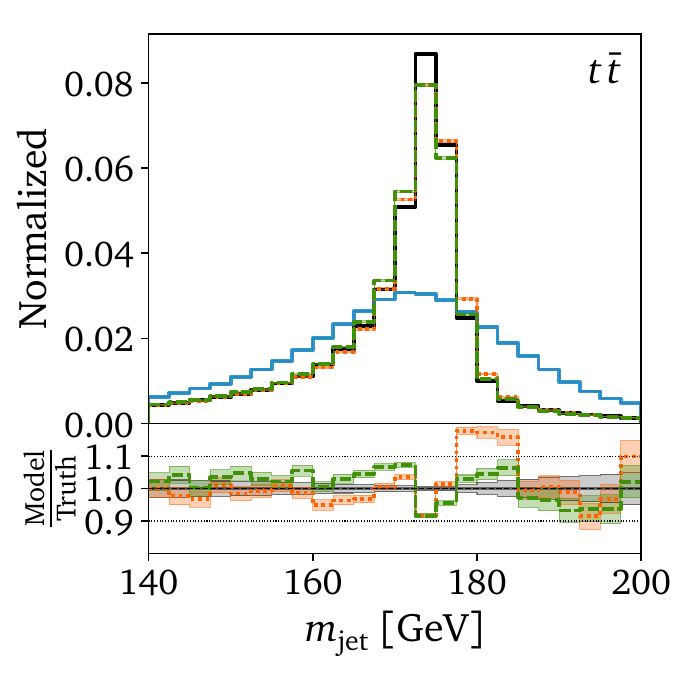}\\
    \includegraphics[width=0.325\textwidth,page=35]{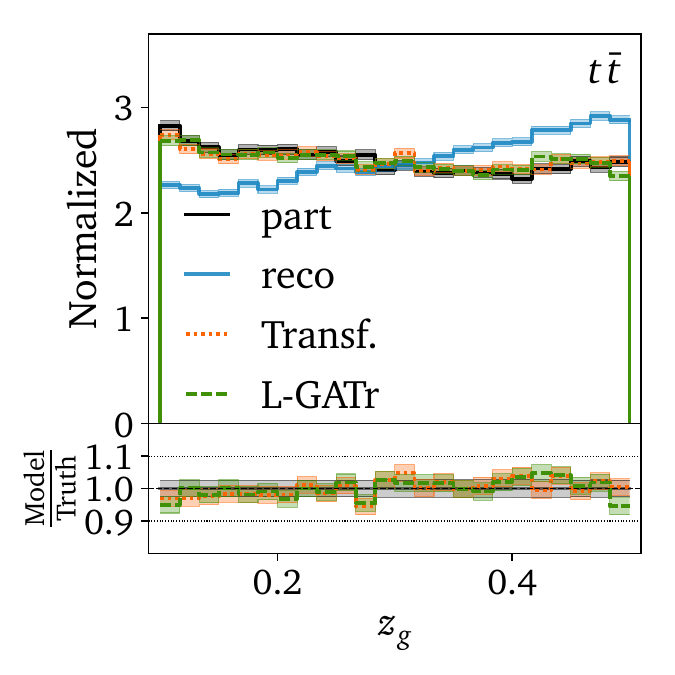}
    \includegraphics[width=0.325\textwidth,page=34]{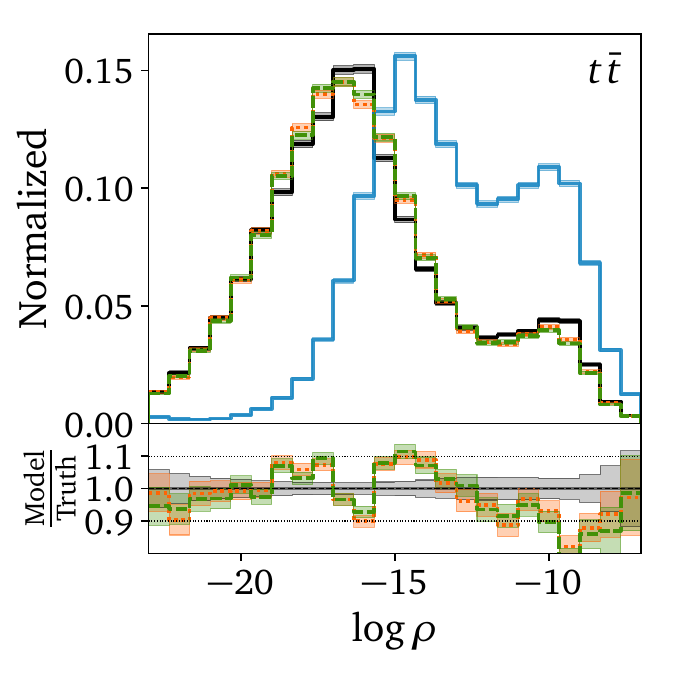}
    \includegraphics[width=0.325\textwidth,page=51]{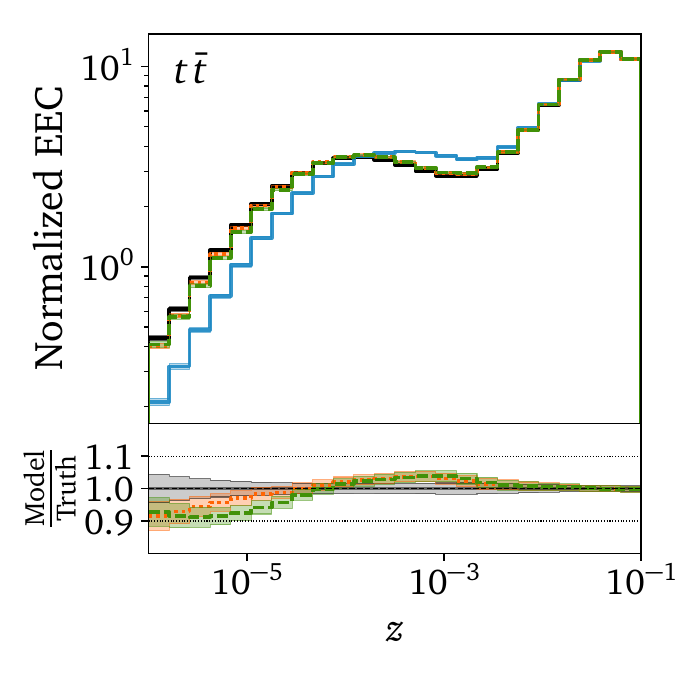} \\
    \includegraphics[width=0.325\textwidth,page=28]{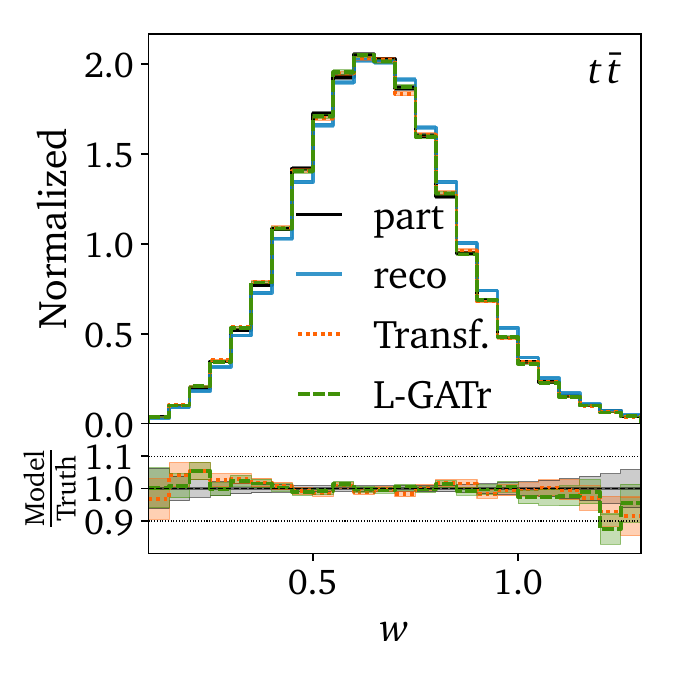}
    \includegraphics[width=0.325\textwidth,page=29]{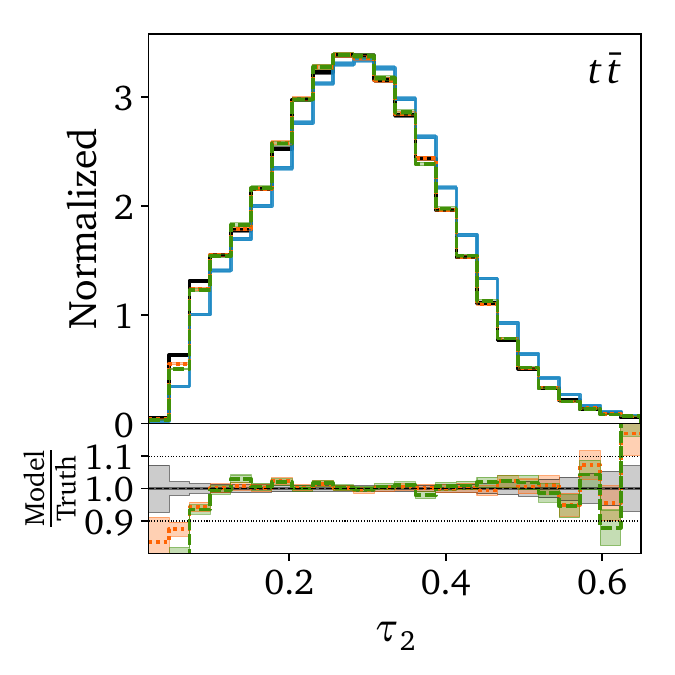}
    \includegraphics[width=0.325\textwidth,page=31]{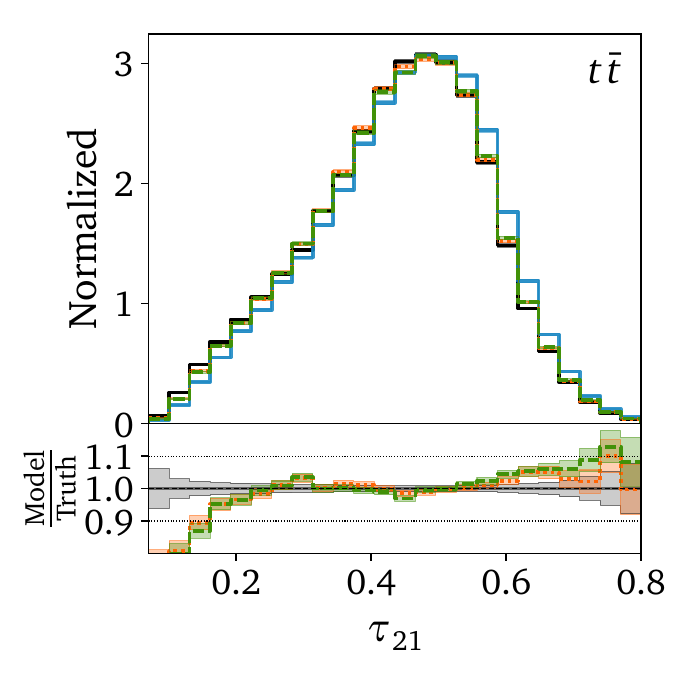}\\
    \includegraphics[width=0.325\textwidth,page=30]{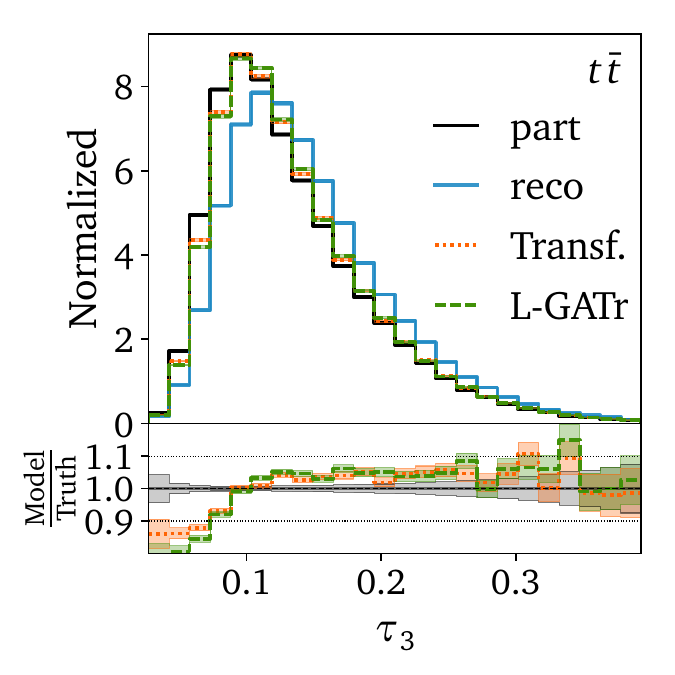}
    \includegraphics[width=0.325\textwidth,page=32]{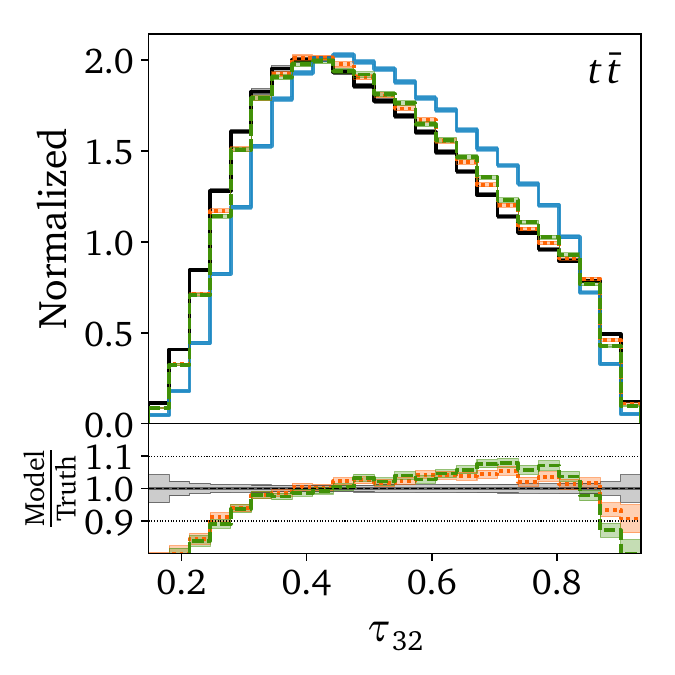}
    \includegraphics[width=0.325\textwidth,page=33]{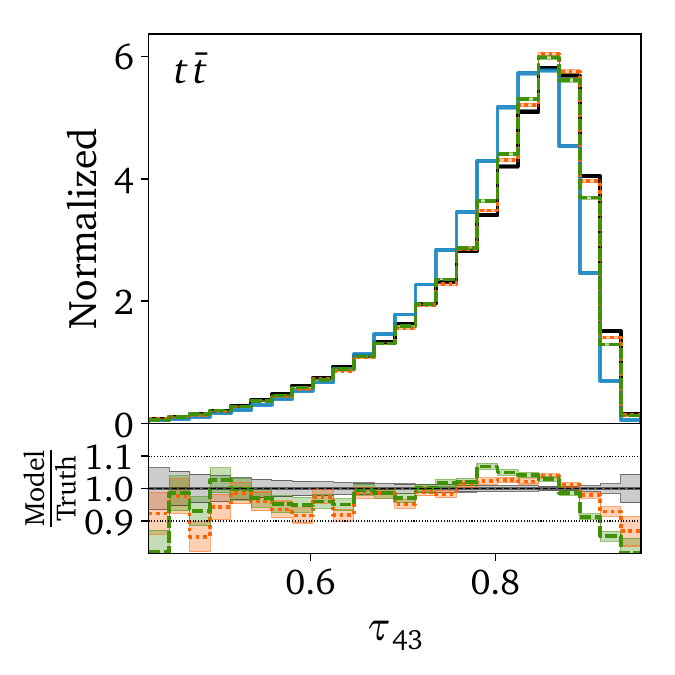}
    \caption{Distributions of jet (substructure) observables computed from the unfolded constituents, directly comparable to the $Z$+jets results in Figures~\ref{fig:z_constituents1} and ~\ref{fig:z_constituents2}.}
    \label{fig:t_constituents}
\end{figure}

First, the lower dimensional tasks of unfolding the constituent multiplicity and jet kinematics are considered. Histograms comparing the detector-level, truth particle-level, and unfolded densities are shown in Figure~\ref{fig:t_mult_jet}. Despite the more complex process, deviations between unfolded and truth distribution remain at the level of few percent in the bulk and exceed 10\% only in the small training statistics tail of the multiplicity distributions. The transformer and L-GATr achieve equal performance with the exception of the invariant mass distribution. Here the simple transformer creates a small deviation in the peak whereas L-GATr reproduces the mass peak perfectly.

The upper panels in Figure~\ref{fig:t_constituents} show the same jet kinematics observables built from the unfolded constituents as Figures~\ref{fig:z_constituents1} and~\ref{fig:z_constituents2}. The main difference between the light-jet case and the top jets is that now the transformer architectures can make a small, but significant difference. Percent level accuracy is observed for the jet transverse momentum and pseudo-rapidity. The jet mass constructed from the constituents now undershoots the mass peak for the standard transformer, while overpopulating the shoulder of the mass peak towards larger momenta. L-GATr performs significantly better on the reconstructed jet mass peak.

At the constituent level, the agreement of the groomed momentum fraction is at the percent level, while for the soft drop mass the challenge in reproducing the narrow peak(s) remains. The EEC shows a slight mismodeling in the tail of the peaked distribution. The shoulder around $z \approx 10^{-3}$ is reproduced reasonably well by both transformer architectures, but deviations appear in the low-statistics tail at $z \lesssim 10^{-4}$. 

Finally, the agreement for the $n$-subjettiness observables in the lower two rows  improves for the top jets over the light jets and is now at percent level. The accuracy in the jet width and $\tau_2$ is much better than for $Z$ plus jets production. For top jets with three-prong substructure, the 3-subjettiness peaks at low values, leading to a challenge for the unfolding network in the steep left shoulder. Similarly, the ratio $\tau_{32}$ is unfolded accurately around the peak, and significant deviations only start to appear for $\tau_{32} \lesssim 0.3$.  The ratio $\tau_{43}$ is unfolded with similar accuracy as $\tau_{21}$, indicating that the generative unfolding consistently reproduces constituent-level correlations.

\clearpage
\section{Conclusion}
\label{sec:conclusion}

This paper presents a novel method for unfolding the full substructure of jets produced in particle collisions, a high- and variable-dimensional  task. While classifier-based methods like OmniFold have shown accuracy for substructure unfolding, this is the first generative unfolding model to accurately unfold the full jet substructure. Complementing the existing classifier-based methods with generative models for accurate high- and variable-dimensional unfolding is important because the two methods can mitigate their respective weaknesses.

The new generative unfolding model employs a conditional 3-step unfolding of the constituent multiplicity, the jet kinematics, and the constituent kinematics. To accommodate a large phase space dimensionality with up to roughly 100 constituents, it employs equivariant transformers (L-GATr) inside a conditional generative CFM network. The conditional setup avoids the subjet-level curse of dimensionality when unfolding the jet-level kinematics to provide the best possible results for analyses focusing on jet-level observables.

The performance of the new generative model was illustrated on two applications, unfolding the substructure of light quark- and gluon-initiated jets and of boosted top jets. The performance of a standard and equivariant transformers were found to be comparable, with slight advantages of L-GATr for the higher-multiplicity top jets. All jet observables, such as the transverse momentum, pseudo-rapidity, and width are unfolded with percent level precision for both light and top jets. 

For the jet constituents, the groomed momentum fraction, the soft drop mass, and an example energy-energy correlator observable are also unfolded at the percent level. The remaining challenge is the 21-subjettiness ratio for light jets. For $Z$ plus jet production, the standard transformer and L-GATr reproduce the subjet kinematics with similar accuracy, with the exception of this one correlation. For the more challenging top jets, the L-GATr unfolding is slightly more accurate than the standard transformer, reflecting the relevance of invariant masses for fat jets. For the top jets, the new generative unfolding reproduces jet-level and subjet observables with high accuracy and without a visible failure mode.

The performance shown in this study marks the state of the art for generative unfolding. The main limitation of this study has been the size of the transformer networks, their training time, and the size of the training dataset. For classifier-based unfolding network ensembles have been shown to lead to significant performance increase. As shown for L-GATr jet tagging~\cite{Brehmer:2024yqw}, larger pre-trained transformers should lead to further improvements of the accuracy.

\subsection*{Code availability}

The code for this paper is available as part of the public Heidelberg hep-ml code and tutorial library \url{https://github.com/heidelberg-hepml/high-dim-unfolding}. The $Z+\text{jets}$ dataset is available on Zenodo \url{https://zenodo.org/records/10668638}, and the $t\bar t$ dataset is available upon request.

\section*{Acknowledgments}

This project has received funding from the European Union’s Horizon Europe research and innovation programme under the Marie Sklodowska-Curie grant agreement No~101168829, \textsl{Challenging AI with Challenges from Physics: How to solve fundamental problems in Physics by AI and vice versa} (AIPHY). AB and SPS acknowledge support by the BMBF Junior Group \textsl{Generative Precision Networks for Particle Physics} (DLR 01IS22079).
JS and TP are supported by the Deutsche Forschungsgemeinschaft (DFG, German Research Foundation) under grant 396021762 – TRR 257 \textsl{Particle Physics Phenomenology after the Higgs Discovery}. JS is funded by the Carl-Zeiss-Stiftung through the project \textsl{Model-Based AI: Physical Models and Deep Learning for Imaging and Cancer Treatment}.  This work is supported by Deutsche Forschungsgemeinschaft (DFG) under Germany’s Excellence Strategy EXC-2181/1 - 390900948 (the Heidelberg STRUCTURES Excellence Cluster).
The authors acknowledge support by the state of Baden-W\"urttemberg through bwHPC and the German Research Foundation (DFG) through the grants INST 35/1597-1 FUGG and INST 39/1232-1 FUGG.   DW and KG are funded by the United States Department of Energy Office of Science.

\appendix

\section{Network and training details}
\label{app:hyperparams}

The architecture and training hyperparameters are listed in Tab.~\ref{tab:z_hyperparameters} for \ref{sec:z_plus_jets} results and in Tab.~\ref{tab:t_hyperparameters} for \ref{sec:ttbar} results. Training times are given for a Nvidia H100 GPU.

\begin{table}[h!]
    \centering
    \begin{footnotesize}
    \begin{tabular}{l|cccccc}
        \toprule
        & \multicolumn{2}{c}{Multiplicity} & \multicolumn{2}{c}{Jet} & \multicolumn{2}{c}{Constituents} \\ 
        & Transformer & L-GATr & Transformer & L-GATr & Transformer & L-GATr \\
        \midrule
        Blocks & 2 encoder & 2 encoder & \makecell{2 encoder \\ 4 decoder} & \makecell{2 encoder \\ 4 decoder} & \makecell{3 encoder \\ 4 decoder} & \makecell{3 encoder \\ 4 decoder} \\
        Channels & 64 & \makecell{8 multivector \\ 8 scalar} & 64 & \makecell{8 multivector \\ 32 scalar} & 128 & \makecell{16 multivector \\ 64 scalar} \\
        Activation & GELU & GELU & GELU & GELU & GELU & GELU \\
        Parameters & $6.8 \cdot 10^4$ & $4.2\cdot 10^4$ & $7.4\cdot 10^5$ & $4.1\cdot 10^5$ & $3.1\cdot 10^6$ & $9.9\cdot 10^5$ \\
        \midrule
        Iterations & $10^6$ & $10^6$ & $2\cdot 10^6$ & $2\cdot 10^6$ & $2\cdot 10^6$ & $2\cdot 10^6$ \\ 
        Batch size & 1024 & 1024 & 1024 & 1024 & 1024 & 1024 \\
        Optimizer & Adam & Adam & Lion & Lion & Lion & Lion \\
        Learning rate & $10^{-3}$ & $10^{-3}$ & $2\cdot10^{-4}$ & $2\cdot10^{-4}$ & $2\cdot10^{-4}$ & $2\cdot10^{-4}$ \\
        Schedule & \makecell{Cosine\\Annealing} & \makecell{Cosine\\Annealing} & \makecell{Cosine\\Annealing} & \makecell{Cosine\\Annealing} & \makecell{Cosine\\Annealing} & \makecell{Cosine\\Annealing} \\
        \makecell{Training time}  & 1.3h & 2.9h & 4.1h & 8.2h & 4.5h & 18.3h \\
        \bottomrule
    \end{tabular}
    \end{footnotesize}
    \caption{Architecture and training hyperparameters for the transformer and L-GATr networks and the three unfolding steps for the Z+jets dataset.}
    \label{tab:z_hyperparameters}
\end{table}

\begin{table}[h!]
    \centering
    \begin{footnotesize}
    \begin{tabular}{l|cccccc}
        \toprule
        & \multicolumn{2}{c}{Multiplicity} & \multicolumn{2}{c}{Jet} & \multicolumn{2}{c}{Constituents} \\ 
        & Transformer & L-GATr & Transformer & L-GATr & Transformer & L-GATr \\
        \midrule
        Blocks & 2 encoder & 2 encoder & \makecell{2 encoder \\ 4 decoder} & \makecell{2 encoder \\ 4 decoder} & \makecell{6 encoder \\ 8 decoder} & \makecell{6 encoder \\ 8 decoder} \\
        Channels & 64 & \makecell{8 multivector \\ 8 scalar} & 64 & \makecell{8 multivector \\ 32 scalar} & 128 & \makecell{16 multivector \\ 64 scalar} \\
        Activation & GELU & GELU & GELU & GELU & GELU & GELU \\
        Parameters & $6.8 \cdot 10^4$ & $4.2\cdot 10^4$ & $7.4\cdot 10^5$ & $4.1\cdot 10^5$ & $6.1\cdot 10^6$ & $2.0\cdot 10^6$ \\
        \midrule
        Iterations & $10^6$ & $10^6$ & $2\cdot 10^6$ & $2\cdot 10^6$ & $2\cdot 10^6$ & $2\cdot 10^6$ \\ 
        Batch size & 1024 & 1024 & 512 & 512 & 512 & 512 \\
        Optimizer & Adam & Adam & Lion & Lion & Lion & Lion \\
        Learning rate & $10^{-3}$ & $10^{-3}$ & $2\cdot10^{-4}$ & $2\cdot10^{-4}$ & $2\cdot10^{-4}$ & $2\cdot10^{-4}$ \\
        Schedule & \makecell{Cosine\\Annealing} & \makecell{Cosine\\Annealing} & \makecell{Cosine\\Annealing} & \makecell{Cosine\\Annealing} & \makecell{Cosine\\Annealing} & \makecell{Cosine\\Annealing} \\
        \makecell{Training time}  & 1.4h & 3.8h & 4.4h & 8.5h & 7.7h & 47.7h \\
        \bottomrule
    \end{tabular}
    \end{footnotesize}
    \caption{Architecture and training hyperparameters for the transformer and L-GATr networks and the three unfolding steps for the top dataset.}
    \label{tab:t_hyperparameters}
\end{table}

\bibliography{tilman,literature}

@article{Moult:2025nhu,
    author = "Moult, Ian and Zhu, Hua Xing",
    title = "{Energy Correlators: A Journey From Theory to Experiment}",
    eprint = "2506.09119",
    archivePrefix = "arXiv",
    primaryClass = "hep-ph",
    month = "6",
    year = "2025"
}

@article{Holguin:2024tkz,
    author = {Holguin, Jack and Moult, Ian and Pathak, Aditya and Procura, Massimiliano and Sch{\"o}fbeck, Robert and Schwarz, Dennis},
    title = "{Top quark mass extractions from energy correlators: a feasibility study}",
    eprint = "2407.12900",
    archivePrefix = "arXiv",
    primaryClass = "hep-ph",
    reportNumber = "DESY-24-107;UWThPh 2024-14, DESY-24-107, UWThPh 2024-14",
    doi = "10.1007/JHEP04(2025)072",
    journal = "JHEP",
    volume = "04",
    pages = "072",
    year = "2025"
}

@article{CMS:2022kqg,
    author = "Tumasyan, Armen and others",
    collaboration = "CMS",
    title = "{Measurement of the differential $\hbox {t}\overline{\hbox {t}}$ production cross section as a function of the jet mass and extraction of the top quark mass in hadronic decays of boosted top quarks}",
    eprint = "2211.01456",
    archivePrefix = "arXiv",
    primaryClass = "hep-ex",
    reportNumber = "CMS-TOP-21-012, CERN-EP-2022-222",
    doi = "10.1140/epjc/s10052-023-11587-8",
    journal = "Eur. Phys. J. C",
    volume = "83",
    number = "7",
    pages = "560",
    year = "2023"
}

@article{Breso:2024jlt,
    author = "Bres{\'o}, V{\'\i}ctor and Heinrich, Gudrun and Magerya, Vitaly and Olsson, Anton",
    title = "{Interpolating amplitudes}",
    eprint = "2412.09534",
    archivePrefix = "arXiv",
    primaryClass = "hep-ph",
    reportNumber = "CERN-TH-2024-211, KA-TP-23-2024, P3H-24-092",
    month = "12",
    year = "2024"
}

@article{Larkoski:2014wba,
    author = "Larkoski, Andrew J. and Marzani, Simone and Soyez, Gregory and Thaler, Jesse",
    title = "{Soft Drop}",
    eprint = "1402.2657",
    archivePrefix = "arXiv",
    primaryClass = "hep-ph",
    reportNumber = "MIT-CTP-4531, DCPT-14-24, IPPP-14-12",
    doi = "10.1007/JHEP05(2014)146",
    journal = "JHEP",
    volume = "05",
    pages = "146",
    year = "2014"
}

@article{Thaler:2010tr,
    author = "Thaler, Jesse and Van Tilburg, Ken",
    title = "{Identifying Boosted Objects with N-subjettiness}",
    eprint = "1011.2268",
    archivePrefix = "arXiv",
    primaryClass = "hep-ph",
    reportNumber = "MIT-CTP-4191",
    doi = "10.1007/JHEP03(2011)015",
    journal = "JHEP",
    volume = "03",
    pages = "015",
    year = "2011"
}

@article{Gong:2022lye,
    author = "Gong, Shiqi and Meng, Qi and Zhang, Jue and Qu, Huilin and Li, Congqiao and Qian, Sitian and Du, Weitao and Ma, Zhi-Ming and Liu, Tie-Yan",
    title = "{An efficient Lorentz equivariant graph neural network for jet tagging}",
    eprint = "2201.08187",
    archivePrefix = "arXiv",
    primaryClass = "hep-ph",
    doi = "10.1007/JHEP07(2022)030",
    journal = "JHEP",
    volume = "07",
    pages = "030",
    year = "2022"
}

@article{Witkowski:2023xmx,
    author = "Witkowski, Edmund and Whiteson, Daniel",
    title = "{Learning Broken Symmetries with Resimulation and Encouraged Invariance}",
    eprint = "2311.05952",
    archivePrefix = "arXiv",
    primaryClass = "hep-ex",
    month = "11",
    year = "2023"
}

@article{Nabat:2024nce,
    author = "Nabat, Seth and Ghosh, Aishik and Witkowski, Edmund and Kasieczka, Gregor and Whiteson, Daniel",
    title = "{Learning broken symmetries with approximate invariance}",
    eprint = "2412.18773",
    archivePrefix = "arXiv",
    primaryClass = "hep-ph",
    doi = "10.1103/PhysRevD.111.072002",
    journal = "Phys. Rev. D",
    volume = "111",
    number = "7",
    pages = "072002",
    year = "2025"
}

@article{Bogatskiy:2022czk,
    author = "Bogatskiy, Alexander and Hoffman, Timothy and Miller, David W. and Offermann, Jan T.",
    title = "{PELICAN: Permutation Equivariant and Lorentz Invariant or Covariant Aggregator Network for Particle Physics}",
    eprint = "2211.00454",
    archivePrefix = "arXiv",
    primaryClass = "hep-ph",
    month = "11",
    year = "2022"
}

@article{lipman2023flowmatchinggenerativemodeling,
      title={Flow Matching for Generative Modeling}, 
      author={Yaron Lipman and Ricky T. Q. Chen and Heli Ben-Hamu and Maximilian Nickel and Matt Le},
      year={2023},
      eprint={2210.02747},
      archivePrefix={arXiv},
      primaryClass={cs.LG},
      url={https://arxiv.org/abs/2210.02747}, 
}

@article{chen2018neural,
  title={Neural ordinary differential equations},
  author={Chen, Ricky TQ and Rubanova, Yulia and Bettencourt, Jesse and Duvenaud, David K},
  journal={{Advances in Neural Information Processing Systems}},
  volume={31},
  year={2018}
}

@article{albergo2023stochastic,
  author="Albergo, Michael S and Boffi, Nicholas M and Vanden-Eijnden, Eric",
  title="Stochastic interpolants: A unifying framework for flows and diffusions",
  year="2023",
  month = "3",
  eprint = "2303.08797",
  archivePrefix = "arXiv",
  primaryClass = "cs.LG"
}

@article{Datta:2018mwd,
    author = "Datta, Kaustuv and Kar, Deepak and Roy, Debarati",
    title = "{Unfolding with Generative Adversarial Networks}",
    eprint = "1806.00433",
    archivePrefix = "arXiv",
    primaryClass = "physics.data-an",
    month = "6",
    year = "2018"
}

@article{Vandegar:2020yvw,
    author = "Vandegar, Maxime and Kagan, Michael and Wehenkel, Antoine and Louppe, Gilles",
    title = "{Neural Empirical Bayes: Source Distribution Estimation and its Applications to Simulation-Based Inference}",
    eprint = "2011.05836",
    archivePrefix = "arXiv",
    primaryClass = "stat.ML",
    month = "11",
    year = "2020"
}

@article{Howard:2021pos,
    author = "Howard, Jessica N. and Mandt, Stephan and Whiteson, Daniel and Yang, Yibo",
    title = "{Learning to simulate high energy particle collisions from unlabeled data}",
    eprint = "2101.08944",
    archivePrefix = "arXiv",
    primaryClass = "hep-ph",
    doi = "10.1038/s41598-022-10966-7",
    journal = "Sci. Rep.",
    volume = "12",
    pages = "7567",
    year = "2022"
}

@article{Backes:2022sph,
    author = "Backes, Mathias and Butter, Anja and Dunford, Monica and Malaescu, Bogdan",
    title = "{An unfolding method based on conditional invertible neural networks (cINN) using iterative training}",
    eprint = "2212.08674",
    archivePrefix = "arXiv",
    primaryClass = "hep-ph",
    doi = "10.21468/scipostphyscore.7.1.007",
    journal = "SciPost Phys. Core",
    volume = "7",
    number = "1",
    pages = "007",
    year = "2024"
}

@article{Diefenbacher:2023wec,
    author = "Diefenbacher, Sascha and Liu, Guan-Horng and Mikuni, Vinicius and Nachman, Benjamin and Nie, Weili",
    title = {{Improving generative model-based unfolding with Schr{\"o}dinger bridges}},
    eprint = "2308.12351",
    archivePrefix = "arXiv",
    primaryClass = "hep-ph",
    doi = "10.1103/PhysRevD.109.076011",
    journal = "Phys. Rev. D",
    volume = "109",
    number = "7",
    pages = "076011",
    year = "2024"
}

@article{Shmakov:2023kjj,
    author = "Shmakov, Alexander and Greif, Kevin and Fenton, Michael and Ghosh, Aishik and Baldi, Pierre and Whiteson, Daniel",
    title = "{End-To-End Latent Variational Diffusion Models for Inverse Problems in High Energy Physics}",
    eprint = "2305.10399",
    archivePrefix = "arXiv",
    primaryClass = "hep-ex",
    month = "5",
    year = "2023"
}

@article{Andreassen:2019cjw,
    author = "Andreassen, Anders and Komiske, Patrick T. and Metodiev, Eric M. and Nachman, Benjamin and Thaler, Jesse",
    title = "{OmniFold: A Method to Simultaneously Unfold All Observables}",
    eprint = "1911.09107",
    archivePrefix = "arXiv",
    primaryClass = "hep-ph",
    reportNumber = "MIT-CTP 5155",
    doi = "10.1103/PhysRevLett.124.182001",
    journal = "Phys. Rev. Lett.",
    volume = "124",
    number = "18",
    pages = "182001",
    year = "2020"
}

@article{Desai:2024yft,
    author = "Desai, Krish and Nachman, Benjamin and Thaler, Jesse",
    title = "{Moment extraction using an unfolding protocol without binning}",
    eprint = "2407.11284",
    archivePrefix = "arXiv",
    primaryClass = "hep-ph",
    reportNumber = "MIT-CTP 5727",
    doi = "10.1103/PhysRevD.110.116013",
    journal = "Phys. Rev. D",
    volume = "110",
    number = "11",
    pages = "116013",
    year = "2024"
}

@article{Falcao:2025jom,
    author = "Falc{\~a}o, Alexandre and Takacs, Adam",
    title = "{High-Dimensional Unfolding in Large Backgrounds}",
    eprint = "2507.06291",
    archivePrefix = "arXiv",
    primaryClass = "hep-ph",
    month = "7",
    year = "2025"
}

@article{Pazos:2024nfe,
    author = "Pazos, Camila and Aeron, Shuchin and Beauchemin, Pierre-Hugues and Croft, Vincent and Huan, Zhengyan and Klassen, Martin and Wongjirad, Taritree",
    title = "{Towards Universal Unfolding of Detector Effects in High-Energy Physics using Denoising Diffusion Probabilistic Models}",
    eprint = "2406.01507",
    archivePrefix = "arXiv",
    primaryClass = "physics.data-an",
    month = "6",
    year = "2024"
}

@article{Shmakov:2024gkd,
    author = "Shmakov, Alexander and Greif, Kevin and Fenton, Michael James and Ghosh, Aishik and Baldi, Pierre and Whiteson, Daniel",
    title = "{Full Event Particle-Level Unfolding with Variable-Length Latent Variational Diffusion}",
    eprint = "2404.14332",
    archivePrefix = "arXiv",
    primaryClass = "hep-ex",
    month = "4",
    year = "2024"
}

@article{Cowan:2002in,
    author = "Cowan, G.",
    editor = "Whalley, M. R. and Lyons, L.",
    title = "{A survey of unfolding methods for particle physics}",
    journal = "Conf. Proc. C",
    volume = "0203181",
    pages = "248--257",
    year = "2002",
    url = "https://www.ippp.dur.ac.uk/Workshops/02/statistics/proceedings//cowan.pdf"
}

@article{Chan:2023tbf,
    author = "Chan, Jay and Nachman, Benjamin",
    title = "{Unbinned profiled unfolding}",
    eprint = "2302.05390",
    archivePrefix = "arXiv",
    primaryClass = "hep-ph",
    doi = "10.1103/PhysRevD.108.016002",
    journal = "Phys. Rev. D",
    volume = "108",
    number = "1",
    pages = "016002",
    year = "2023"
}

@article{Zhu:2024drd,
    author = "Zhu, Huanbiao and Desai, Krish and Kuusela, Mikael and Mikuni, Vinicius and Nachman, Benjamin and Wasserman, Larry",
    title = "{Multidimensional Deconvolution with Profiling}",
    eprint = "2409.10421",
    archivePrefix = "arXiv",
    primaryClass = "hep-ph",
    month = "9",
    year = "2024"
}

@article{H1:2021wkz,
    author = "Andreev, V. and others",
    collaboration = "H1",
    title = "{Measurement of Lepton-Jet Correlation in Deep-Inelastic Scattering with the H1 Detector Using Machine Learning for Unfolding}",
    eprint = "2108.12376",
    archivePrefix = "arXiv",
    primaryClass = "hep-ex",
    reportNumber = "DESY 21-130",
    doi = "10.1103/PhysRevLett.128.132002",
    journal = "Phys. Rev. Lett.",
    volume = "128",
    number = "13",
    pages = "132002",
    year = "2022"
}

@article{H1:2023fzk,
    author = "Andreev, V. and others",
    collaboration = "H1",
    title = "{Unbinned deep learning jet substructure measurement in high Q2ep collisions at HERA}",
    eprint = "2303.13620",
    archivePrefix = "arXiv",
    primaryClass = "hep-ex",
    reportNumber = "DESY-23-034",
    doi = "10.1016/j.physletb.2023.138101",
    journal = "Phys. Lett. B",
    volume = "844",
    pages = "138101",
    year = "2023"
}

@article{H1:2024mox,
    author = "Andreev, V. and others",
    collaboration = "H1",
    title = "{Machine Learning-Assisted Measurement of Lepton-Jet Azimuthal Angular Asymmetries in Deep-Inelastic Scattering at HERA}",
    eprint = "2412.14092",
    archivePrefix = "arXiv",
    primaryClass = "hep-ex",
    reportNumber = "DESY24-200",
    month = "12",
    year = "2024"
}

@article{LHCb:2022rky,
    author = "Aaij, Roel and others",
    collaboration = "LHCb",
    title = "{Multidifferential study of identified charged hadron distributions in $Z$-tagged jets in proton-proton collisions at $\sqrt{s}=$13 TeV}",
    eprint = "2208.11691",
    archivePrefix = "arXiv",
    primaryClass = "hep-ex",
    reportNumber = "CERN-EP-2022-161, LHCb-PAPER-2022-013",
    doi = "10.1103/PhysRevD.108.L031103",
    journal = "Phys. Rev. D",
    volume = "108",
    pages = "L031103",
    year = "2023"
}

@article{Song:2023sxb,
    author = "Song, Youqi",
    collaboration = "STAR",
    title = "{Measurement of CollinearDrop jet mass and its correlation with SoftDrop groomed jet substructure observables in $\sqrt{s}=200$ GeV $pp$ collisions by STAR}",
    eprint = "2307.07718",
    archivePrefix = "arXiv",
    primaryClass = "nucl-ex",
    month = "7",
    year = "2023"
}

@article{Pani:2024mgy,
    author = "Pani, Tanmay",
    collaboration = "STAR",
    title = "{Generalized angularities measurements from STAR at {\ensuremath{\sqrt{}}}SNN = 200 GeV}",
    eprint = "2403.13921",
    archivePrefix = "arXiv",
    primaryClass = "nucl-ex",
    doi = "10.1051/epjconf/202429611003",
    journal = "EPJ Web Conf.",
    volume = "296",
    pages = "11003",
    year = "2024"
}

@article{Huang:2025ziq,
    author = "Huang, Roger G. and Cudd, Andrew and Kawaue, Masaki and Kikawa, Tatsuya and Nachman, Benjamin and Mikuni, Vinicius and Wilkinson, Callum",
    title = "{Machine learning assisted unfolding for neutrino cross-section measurements with the OmniFold technique}",
    eprint = "2504.06857",
    archivePrefix = "arXiv",
    primaryClass = "physics.data-an",
    doi = "10.1103/sp1f-n9k2",
    journal = "Phys. Rev. D",
    volume = "112",
    number = "1",
    pages = "012008",
    year = "2025"
}

@article{ATLAS:2025qtv,
    author = "Aad, Georges and others",
    collaboration = "ATLAS",
    title = "{Measurement of jet track functions in pp collisions at s=13 TeV with the ATLAS detector}",
    eprint = "2502.02062",
    archivePrefix = "arXiv",
    primaryClass = "hep-ex",
    reportNumber = "CERN-EP-2024-333",
    doi = "10.1016/j.physletb.2025.139680",
    journal = "Phys. Lett. B",
    volume = "868",
    pages = "139680",
    year = "2025"
}

@article{CMS:2025sws,
    author = "Chekhovsky, Vladimir and others",
    collaboration = "CMS",
    title = "{Measurement of event shapes in minimum-bias events from proton-proton collisions at $\sqrt{s}$ = 13 TeV}",
    eprint = "2505.17850",
    archivePrefix = "arXiv",
    primaryClass = "hep-ex",
    reportNumber = "CMS-SMP-23-008, CERN-EP-2025-041",
    month = "5",
    year = "2025"
}

@article{Canelli:2025ybb,
    author = "Canelli, Florencia and others",
    title = "{A Practical Guide to Unbinned Unfolding}",
    eprint = "2507.09582",
    archivePrefix = "arXiv",
    primaryClass = "hep-ph",
    month = "7",
    year = "2025"
}

@article{Komiske:2022vxg,
    author = "Komiske, Patrick T. and Kryhin, Serhii and Thaler, Jesse",
    title = "{Disentangling quarks and gluons in CMS open data}",
    eprint = "2205.04459",
    archivePrefix = "arXiv",
    primaryClass = "hep-ph",
    reportNumber = "MIT-CTP 5422",
    doi = "10.1103/PhysRevD.106.094021",
    journal = "Phys. Rev. D",
    volume = "106",
    number = "9",
    pages = "094021",
    year = "2022"
}

@article{Lee:2022uwt,
    author = "Lee, Kyle and Me{\c{c}}aj, Bianka and Moult, Ian",
    title = "{Conformal collider physics meets LHC data}",
    eprint = "2205.03414",
    archivePrefix = "arXiv",
    primaryClass = "hep-ph",
    doi = "10.1103/PhysRevD.111.L011502",
    journal = "Phys. Rev. D",
    volume = "111",
    number = "1",
    pages = "L011502",
    year = "2025"
}

@article{Desai:2025mpy,
    author = "Desai, Krish and Long, Owen and Nachman, Benjamin",
    title = "{Unbinned Inference with Correlated Events}",
    eprint = "2504.14072",
    archivePrefix = "arXiv",
    primaryClass = "physics.data-an",
    month = "4",
    year = "2025"
}

@article{Cacciari:2008gp,
    author = "Cacciari, Matteo and Salam, Gavin P. and Soyez, Gregory",
    title = "{The anti-$k_t$ jet clustering algorithm}",
    eprint = "0802.1189",
    archivePrefix = "arXiv",
    primaryClass = "hep-ph",
    reportNumber = "LPTHE-07-03",
    doi = "10.1088/1126-6708/2008/04/063",
    journal = "JHEP",
    volume = "04",
    pages = "063",
    year = "2008"
}

@article{deFavereau:2013fsa,
    author = "de Favereau, J. and Delaere, C. and Demin, P. and Giammanco, A. and Lema{\^\i}tre, V. and Mertens, A. and Selvaggi, M.",
    collaboration = "DELPHES 3",
    title = "{DELPHES 3, A modular framework for fast simulation of a generic collider experiment}",
    eprint = "1307.6346",
    archivePrefix = "arXiv",
    primaryClass = "hep-ex",
    doi = "10.1007/JHEP02(2014)057",
    journal = "JHEP",
    volume = "02",
    pages = "057",
    year = "2014"
}

@article{Cacciari:2011ma,
    author = "Cacciari, Matteo and Salam, Gavin P. and Soyez, Gregory",
    title = "{FastJet User Manual}",
    eprint = "1111.6097",
    archivePrefix = "arXiv",
    primaryClass = "hep-ph",
    reportNumber = "CERN-PH-TH-2011-297",
    doi = "10.1140/epjc/s10052-012-1896-2",
    journal = "Eur. Phys. J. C",
    volume = "72",
    pages = "1896",
    year = "2012"
}

@article{Alwall:2011uj,
    author = "Alwall, Johan and Herquet, Michel and Maltoni, Fabio and Mattelaer, Olivier and Stelzer, Tim",
    title = "{MadGraph 5 : Going Beyond}",
    eprint = "1106.0522",
    archivePrefix = "arXiv",
    primaryClass = "hep-ph",
    reportNumber = "FERMILAB-PUB-11-448-T",
    doi = "10.1007/JHEP06(2011)128",
    journal = "JHEP",
    volume = "06",
    pages = "128",
    year = "2011"
}

@article{Sjostrand:2014zea,
    author = {Sj{\"o}strand, Torbj{\"o}rn and Ask, Stefan and Christiansen, Jesper R. and Corke, Richard and Desai, Nishita and Ilten, Philip and Mrenna, Stephen and Prestel, Stefan and Rasmussen, Christine O. and Skands, Peter Z.},
    title = "{An introduction to PYTHIA 8.2}",
    eprint = "1410.3012",
    archivePrefix = "arXiv",
    primaryClass = "hep-ph",
    reportNumber = "LU-TP-14-36, MCNET-14-22, CERN-PH-TH-2014-190, FERMILAB-PUB-14-316-CD, DESY-14-178, SLAC-PUB-16122",
    doi = "10.1016/j.cpc.2015.01.024",
    journal = "Comput. Phys. Commun.",
    volume = "191",
    pages = "159--177",
    year = "2015"
}

@article{Butter:2025mek,
    author = "Butter, Anja and Huetsch, Nathan and Mikuni, Vinicius and Nachman, Benjamin and Palacios Schweitzer, Sofia",
    title = "{Analysis-ready Generative Unfolding}",
    eprint = "2509.02708",
    archivePrefix = "arXiv",
    primaryClass = "hep-ph",
    month = "9",
    year = "2025"
}

@article{Badea:2025wzd,
    author = "Badea, Anthony and others",
    title = "{Analysis note: measurement of thrust in $e^{+}e^{-}$ collisions at $\sqrt{s}$ = 91 GeV with archived ALEPH data}",
    eprint = "2507.14349",
    archivePrefix = "arXiv",
    primaryClass = "hep-ex",
    month = "7",
    year = "2025"
}

@article{Favaro:2025pgz,
    author = "Favaro, Luigi and Gerhartz, Gerrit and Hamprecht, Fred A. and Lippmann, Peter and Pitz, Sebastian and Plehn, Tilman and Qu, Huilin and Spinner, Jonas",
    title = "{Lorentz-Equivariance without Limitations}",
    eprint = "2508.14898",
    archivePrefix = "arXiv",
    primaryClass = "hep-ph",
    month = "8",
    year = "2025"
}

@article{Butter:2025via,
    author = "Butter, Anja and Heimel, Theo and Huetsch, Nathan and Kagan, Michael and Plehn, Tilman",
    title = "{Simulation-Prior Independent Neural Unfolding Procedure}",
    eprint = "2507.15084",
    archivePrefix = "arXiv",
    primaryClass = "hep-ph",
    month = "7",
    year = "2025"
}

@article{Favaro:2025psi,
    author = "Favaro, Luigi and Kogler, Roman and Paasch, Alexander and Palacios Schweitzer, Sofia and Plehn, Tilman and Schwarz, Dennis",
    title = "{How to Unfold Top Decays}",
    eprint = "2501.12363",
    archivePrefix = "arXiv",
    primaryClass = "hep-ph",
    reportNumber = "DESY-25-012",
    doi = "10.21468/SciPostPhysCore.8.3.053",
    journal = "SciPost Phys. Core",
    volume = "8",
    pages = "053",
    year = "2025"
}

@article{Bahl:2024gyt,
    author = "Bahl, Henning and Elmer, Nina and Favaro, Luigi and Hau{\ss}mann, Manuel and Plehn, Tilman and Winterhalder, Ramon",
    title = "{Accurate Surrogate Amplitudes with Calibrated Uncertainties}",
    eprint = "2412.12069",
    archivePrefix = "arXiv",
    primaryClass = "hep-ph",
    month = "12",
    year = "2024"
}

@article{ATLAS:2024rpl,
    author = "Aad, Georges and others",
    collaboration = "ATLAS",
    title = "{Precision calibration of calorimeter signals in the ATLAS experiment using an uncertainty-aware neural network}",
    eprint = "2412.04370",
    archivePrefix = "arXiv",
    primaryClass = "hep-ex",
    reportNumber = "CERN-EP-2024-317",
    month = "12",
    year = "2024"
}

@article{Butter:2024vbx,
    author = "Butter, Anja and Diefenbacher, Sascha and Huetsch, Nathan and Mikuni, Vinicius and Nachman, Benjamin and Palacios Schweitzer, Sofia and Plehn, Tilman",
    title = "{Generative unfolding with distribution mapping}",
    eprint = "2411.02495",
    archivePrefix = "arXiv",
    primaryClass = "hep-ph",
    doi = "10.21468/SciPostPhys.18.6.200",
    journal = "SciPost Phys.",
    volume = "18",
    number = "6",
    pages = "200",
    year = "2025"
}

@article{Brehmer:2024yqw,
    author = "Brehmer, Johann and Bres{\'o}, V{\'\i}ctor and de Haan, Pim and Plehn, Tilman and Qu, Huilin and Spinner, Jonas and Thaler, Jesse",
    title = "{A Lorentz-Equivariant Transformer for All of the LHC}",
    eprint = "2411.00446",
    archivePrefix = "arXiv",
    primaryClass = "hep-ph",
    reportNumber = "MIT-CTP/5802",
    month = "11",
    year = "2024"
}

@article{Huetsch:2024quz,
    author = "Huetsch, Nathan and others",
    title = "{The landscape of unfolding with machine learning}",
    eprint = "2404.18807",
    archivePrefix = "arXiv",
    primaryClass = "hep-ph",
    doi = "10.21468/SciPostPhys.18.2.070",
    journal = "SciPost Phys.",
    volume = "18",
    number = "2",
    pages = "070",
    year = "2025"
}

@article{Heimel:2023mvw,
    author = "Heimel, Theo and Huetsch, Nathan and Winterhalder, Ramon and Plehn, Tilman and Butter, Anja",
    title = "{Precision-machine learning for the matrix element method}",
    eprint = "2310.07752",
    archivePrefix = "arXiv",
    primaryClass = "hep-ph",
    reportNumber = "IRMP-CP3-23-55",
    doi = "10.21468/SciPostPhys.17.5.129",
    journal = "SciPost Phys.",
    volume = "17",
    number = "5",
    pages = "129",
    year = "2024"
}

@article{Ackerschott:2023nax,
    author = "Ackerschott, Jona and Barman, Rahool Kumar and Gon{\c{c}}alves, Dorival and Heimel, Theo and Plehn, Tilman",
    title = "{Returning CP-observables to the frames they belong}",
    eprint = "2308.00027",
    archivePrefix = "arXiv",
    primaryClass = "hep-ph",
    doi = "10.21468/SciPostPhys.17.1.001",
    journal = "SciPost Phys.",
    volume = "17",
    number = "1",
    pages = "001",
    year = "2024"
}

@article{Butter:2023fov,
    author = "Butter, Anja and Huetsch, Nathan and Palacios Schweitzer, Sofia and Plehn, Tilman and Sorrenson, Peter and Spinner, Jonas",
    title = "{Jet diffusion versus JetGPT {\textendash} Modern networks for the LHC}",
    eprint = "2305.10475",
    archivePrefix = "arXiv",
    primaryClass = "hep-ph",
    doi = "10.21468/SciPostPhysCore.8.1.026",
    journal = "SciPost Phys. Core",
    volume = "8",
    pages = "026",
    year = "2025"
}

@article{Plehn:2022ftl,
    author = "Plehn, Tilman and Butter, Anja and Dillon, Barry and Heimel, Theo and Krause, Claudius and Winterhalder, Ramon",
    title = "{Modern Machine Learning for LHC Physicists}",
    eprint = "2211.01421",
    archivePrefix = "arXiv",
    primaryClass = "hep-ph",
    month = "11",
    year = "2022"
}

@article{Bellagente:2020piv,
    author = {Bellagente, Marco and Butter, Anja and Kasieczka, Gregor and Plehn, Tilman and Rousselot, Armand and Winterhalder, Ramon and Ardizzone, Lynton and K{\"o}the, Ullrich},
    title = "{Invertible Networks or Partons to Detector and Back Again}",
    eprint = "2006.06685",
    archivePrefix = "arXiv",
    primaryClass = "hep-ph",
    doi = "10.21468/SciPostPhys.9.5.074",
    journal = "SciPost Phys.",
    volume = "9",
    pages = "074",
    year = "2020"
}

@article{Bellagente:2019uyp,
    author = "Bellagente, Marco and Butter, Anja and Kasieczka, Gregor and Plehn, Tilman and Winterhalder, Ramon",
    title = "{How to GAN away Detector Effects}",
    eprint = "1912.00477",
    archivePrefix = "arXiv",
    primaryClass = "hep-ph",
    doi = "10.21468/SciPostPhys.8.4.070",
    journal = "SciPost Phys.",
    volume = "8",
    number = "4",
    pages = "070",
    year = "2020"
}

@article{Plehn:2011tg,
    author = "Plehn, Tilman and Spannowsky, Michael",
    title = "{Top Tagging}",
    eprint = "1112.4441",
    archivePrefix = "arXiv",
    primaryClass = "hep-ph",
    doi = "10.1088/0954-3899/39/8/083001",
    journal = "J. Phys. G",
    volume = "39",
    pages = "083001",
    year = "2012"
}

@article{Plehn:2009rk,
    author = "Plehn, Tilman and Salam, Gavin P. and Spannowsky, Michael",
    title = "{Fat Jets for a Light Higgs}",
    eprint = "0910.5472",
    archivePrefix = "arXiv",
    primaryClass = "hep-ph",
    reportNumber = "KA-TP-12-2009",
    doi = "10.1103/PhysRevLett.104.111801",
    journal = "Phys. Rev. Lett.",
    volume = "104",
    pages = "111801",
    year = "2010"
}
\end{document}